\documentclass[traditabstract]{aa} 
\usepackage{savesym}
\usepackage{amsmath}
\savesymbol{iint}
\usepackage{graphicx}
\usepackage{xspace}
\usepackage{booktabs}
\usepackage[varg]{txfonts}
\usepackage{natbib,twoopt}
\usepackage[breaklinks=f]{hyperref}
\usepackage{epsfig}
\usepackage{txfonts}
\usepackage{amssymb}
\usepackage{hyperref}
\usepackage{color}
\restoresymbol{TXF}{iint}
\bibpunct{(}{)}{;}{a}{}{,} 
 
\newcommandtwoopt{\citeads}[3][][]{\href{http://adsabs.harvard.edu/abs/#3}%
{\citealp[#1][#2]{#3}}}
\newcommandtwoopt{\citepads}[3][][]{\href{http://adsabs.harvard.edu/abs/#3}%
{\citep[#1][#2]{#3}}}
\newcommandtwoopt{\citetads}[3][][]{\href{http://adsabs.harvard.edu/abs/#3}%
{\citet[#1][#2]{#3}}} 
\newcommandtwoopt{\citeyearads}[3][][]%
{\href{http://adsabs.harvard.edu/abs/#3}{\citeyear[#1][#2]{#3}}}

\newcommand{\emth}[1]{\ensuremath{#1}\xspace}
\newcommand{\parm}[1]{\ensuremath{#1}\xspace}

\newcommand{\mearth}{\emth{\mathrm{M_{\oplus}}}} 	 

\newcommand{\pfr}{\parm{f}}                               
\newcommand{\paa}{\parm{R_\mathrm{p}^{2}/R_{\star}^{2}}}  
\newcommand{\pim}{\parm{b}}                               
\newcommand{\pper}{\parm{P_\mathrm{orb}}}                 
\newcommand{\ptc}{\parm{T_\mathrm{0}}}                    
\newcommand{\pec}{\parm{e}}                               
\newcommand{\pom}{\parm{\omega}}                          
\newcommand{\srho}{\parm{\rho_\star}}                     

\newcommand{\pelc}{\parm{\sigma_\mathrm{LC}}}             
\newcommand{\pesc}{\parm{\sigma_\mathrm{SC}}}             

\newcommand{\ssc}{\ensuremath{\mathrm{SC}}\xspace}
\newcommand{\slc}{\ensuremath{\mathrm{LC}}\xspace}

\newcommand{\ecl}{\ensuremath{\mathrm{EC}}\xspace}

\newcommand{\pv}{\ensuremath{\vec{\theta}}\xspace}

 \def\teff{$T_\mathrm{eff}$}                 
 \def\m2s2{\hbox{\,m$^{2}$\,s$^{-2}$}}       
 \def\kms{\hbox{\,km\,s$^{-1}$}}             
 \def\cms2{\hbox{\,cm\,s$^{-2}$}}           
 \def\gcm3{\hbox{\,g\,cm$^{-3}$}}            
 \def\vsini{\hbox{$v$\,sin\,$i_{\star}$}}    
 \def\sini{\hbox{sin\,$i_{\star}$}}          
 \def\Msun{\hbox{$\mathrm{M}_{\odot}$}}      
 \def\Rsun{\hbox{$\mathrm{R}_{\odot}$}}      
 \def\Mjup{\hbox{$\mathrm{M}_\mathrm{Jup}$}} 
 \def\Rjup{\hbox{$\mathrm{R}_\mathrm{Jup}$}} 
 \def\mp{{\emph M}$_\mathrm{p}$}             
 \def\rp{{\emph R}$_\mathrm{p}$}             
 \def\corot{\emph{CoRoT}}                    
 \def\kepler{\emph{Kepler}}                  

\begin{document}

\title{Kepler-423b: a half-Jupiter mass planet transiting \\ a very old solar-like star\,\thanks{Based on observations obtained with the Nordic Optical Telescope, operated on the island of La Palma jointly by Denmark, Finland, Iceland, Norway, and Sweden, in the Spanish Observatorio del Roque de los Muchachos of the Instituto de Astrofisica de Canarias, in time allocated by OPTICON and the Spanish Time Allocation Committee (CAT).}$^,$\thanks{The research leading to these results has received funding from the European Community's Seventh Framework Programme (FP7/2007-2013) under grant agreement number RG226604 (OPTICON) and 267251 (AstroFIt).}}


   \author{D.~Gandolfi\inst{\ref{LSW},\ref{OACT}}
      \and H.~Parviainen\inst{\ref{Oxford}}
      \and H.\,J.~Deeg\inst{\ref{IAC},\ref{LaLaguna}}
      \and A.\,F.~Lanza\inst{\ref{OACT}}
      \and M.~Fridlund\inst{\ref{DLR},\ref{Onsala},\ref{Leiden}}
      \and P.\,G.~Prada~Moroni\inst{\ref{UniPisa},\ref{INAFPisa}}
      \and R.~Alonso\inst{\ref{IAC},\ref{LaLaguna}}
      \and T.~Augusteijn\inst{\ref{NOT}}
      \and J.~Cabrera\inst{\ref{DLR}}
      \and T.~Evans\inst{\ref{Oxford}}
      \and S.~Geier\inst{\ref{NOT}}
      \and A.\,P.~Hatzes\inst{\ref{Tautenburg}}
      \and T.~Holczer\inst{\ref{TelAviv}}
      \and S.~Hoyer\inst{\ref{IAC},\ref{LaLaguna}}
      \and T.~Kangas\inst{\ref{Turku},\ref{NOT}}
      \and T.~Mazeh\inst{\ref{TelAviv}}
      \and I.~Pagano\inst{\ref{OACT}}
      \and L.~Tal-Or\inst{\ref{TelAviv}}
      \and B.~Tingley\inst{\ref{SAC}}
      }

\institute{Landessternwarte K\"onigstuhl, Zentrum f\"ur Astronomie der Universit\"at Heidelberg, K\"onigstuhl 12, D-69117 Heidelberg, Germany\label{LSW}\\
           \email{davide.gandolfi@lsw.uni-heidelberg.de}
      \and INAF - Osservatorio Astrofisico di Catania, Via S. Sofia 78, 95123 Catania, Italy\label{OACT}
      \and Department of Physics, University of Oxford, Oxford, OX1 3RH, United Kingdom\label{Oxford}
      \and Instituto de Astrof\'\i sica de Canarias, C/\,V\'\i a L\'actea s/n, 38205 La Laguna, Spain\label{IAC}
      \and Departamento de Astrof\'\i sica, Universidad de La Laguna, 38206 La Laguna, Spain\label{LaLaguna}
      \and Institute of Planetary Research, German Aerospace Center, Rutherfordstrasse 2, 12489 Berlin, Germany\label{DLR}
      \and Department of Earth and Space Sciences, Chalmers University of Technology, Onsala Space Observatory, 439 92, Onsala, Sweden\label{Onsala}
      \and Leiden Observatory, University of Leiden, PO Box 9513, 2300 RA, Leiden, The Netherlands\label{Leiden}
      \and Physics Department ``E. Fermi'' University of Pisa, Largo B. Pontecorvo 3, 56127 Pisa, Italy\label{UniPisa}
      \and Istituto Nazionale di Fisica Nucleare, Largo B. Pontecorvo 3, 56127 Pisa, Italy\label{INAFPisa}
      \and Nordic Optical Telescope, Apartado 474, 38700 Santa Cruz de La Palma, Spain\label{NOT}
      \and Th\"uringer Landessternwarte, Sternwarte 5, D-07778 Tautenburg, Germany\label{Tautenburg}
      \and School of Physics and Astronomy, Raymond and Beverly Sackler Faculty of Exact Sciences, Tel Aviv University, Tel Aviv, Israel\label{TelAviv}
      \and Tuorla Observatory, Department of Physics and Astronomy, University of Turku, V\"ais\"al\"antie 20, FI-21500 Piikki\"o, Finland\label{Turku}
      \and Stellar Astrophysics Centre, Department of Physics and Astronomy, $\AA$rhus Uni., Ny Munkegade 120, DK-8000 $\AA$rhus C, Denmark\label{SAC}
           }

\date{Received 26 September 2014 / Accepted 17 November 2014}

 
\abstract{We report the spectroscopic confirmation of the \kepler\ object of interest KOI-183.01 (Kepler-423b), a half-Jupiter mass planet transiting an old solar-like star every 2.7 days. Our analysis is the first to combine the full \kepler\ photometry (quarters 1-17) with high-precision radial velocity measurements taken with the FIES spectrograph at the Nordic Optical Telescope. We simultaneously modelled the photometric and spectroscopic data-sets using Bayesian approach coupled with Markov chain Monte Carlo sampling. We found that the \kepler\ pre-search data conditioned light curve of Kepler-423 exhibits quarter-to-quarter systematic variations of the transit depth, with a peak-to-peak amplitude of $\sim$4.3\,\% and seasonal trends reoccurring every four quarters. We attributed these systematics to an incorrect assessment of the quarterly variation of the crowding metric. The host star Kepler-423 is a G4 dwarf with $M_\star=0.85\pm0.04$\,\Msun, $R_\star=0.95\pm0.04$\,\Rsun, \teff~$=5560\pm80$\,K, [M/H]=$-$0.10$\pm$0.05 dex, and with an age of $11\pm2$\,Gyr. The planet Kepler-423b has a mass of \mp~$=0.595\pm0.081$~\Mjup\ and a radius of \rp~$=1.192\pm0.052$~\Rjup, yielding a planetary bulk density of $\rho_\mathrm{p}=0.459\pm0.083$~\gcm3. The radius of Kepler-423b is consistent with both theoretical models for irradiated coreless giant planets and expectations based on empirical laws. The inclination of the stellar spin axis suggests that the system is aligned along the line of sight. We detected a tentative secondary eclipse of the planet at a 2-$\sigma$ confidence level ($\Delta F_{\mathrm{ec}}=14.2\pm6.6$~ppm) and found that the orbit might have a small non-zero eccentricity of $0.019^{+0.028}_{-0.014}$. With a Bond albedo of $A_\mathrm{B}=0.037\pm0.019$, Kepler-423b is one of the gas-giant planets with the lowest albedo known so far.}

               
\titlerunning{The transiting planet Kepler-423b}
\authorrunning{Gandolfi et al.}

   \maketitle
%

\section{Introduction}
\label{Introduction}

We can rightfully argue that space-based transit surveys such as \corot\ \citep{Baglin2006} and \kepler\ \citep{Borucki2010} have revolutionised the field of exoplanetary science. Their high-precision and nearly uninterrupted photometry has opened up the doors to planet parameter spaces that are not easily accessible from the ground, most notably, the Earth-radius planet domain \citep[e.g.,][] {Leger2009,Sanchis-Ojeda2013,Quintana2014}.

When combined with high-resolution spectroscopy, space-based photometry provides us with the most precise planetary and stellar parameters, which in turn are essential for the investigation of the internal structure, migration, and evolution of the planets \citep{Rauer2014}. The exquisite photometric accuracy allows us to detect the eclipse of hot Jupiters even in the visible \citep[e.g.,][]{Coughlin2012,Parviainen2013}. Measuring the eclipse of transiting exoplanets -- also known as planet occultation, secondary eclipse, and secondary transit -- is a powerful tool for probing their atmospheres, in particular their albedos and brightness temperatures \citep{Winn2010a}. The timing and duration of the secondary eclipse, coupled with the timing and duration of the transit, enable us to measure small non-zero eccentricities ($e\lesssim0.1$) that are not easily detectable with radial velocity (RV) measurements. The eccentricity is an important parameter for the investigation of the star-planet tidal interactions, planet-planet gravitational perturbations, and migration mechanisms of hot Jupiters.

\begin{table}[t]
\label{StarTable}      
  \caption{KIC, KOI, GSC2.3, USNO-A2, and 2MASS identifiers of the planet-hosting star Kepler-423. 
           Equatorial coordinates and optical SDSS-$g$,-$r$,-$i$,-$z$ photometry are from
           the \kepler\ Input Catalogue. Infrared $J$,$H$,$Ks$ and $W1$,$W2$,$W3$,$W4$ data are
           from the 2MASS \citep{Cutri2003} and WISE All-Sky Data Release \citep{Wright2010,Cutri2012} 
           database, respectively.}
  \centering
  \begin{tabular}{lll}       
  \multicolumn{1}{l}{\emph{Main identifiers}}     \\
  \hline
  \hline
  \noalign{\smallskip}                
  KIC             & 9651668           \\
  KOI~ID          & 183               \\
  GSC2.3~ID       & N2JG036249        \\
  USNO-A2~ID      & 1350-10669726     \\
  2MASS~ID        & 19312537+4623282  \\
  \noalign{\smallskip}                
  \hline
  \noalign{\medskip}
  \noalign{\smallskip}                
  \multicolumn{2}{l}{\emph{Equatorial coordinates}}     \\
  \hline                                  
  \hline                                  
  \noalign{\smallskip}                
  RA \,(J2000)      & $19^h\,31^m\,25\fs378$              \\
  Dec (J2000)       & $+46\degr\,23\arcmin\,28\farcs240$  \\
  \noalign{\smallskip}                
  \hline
  \noalign{\medskip}
  \noalign{\smallskip}                
  \multicolumn{3}{l}{\emph{Magnitudes}} \\
  \hline
  \hline
  \noalign{\smallskip}                
  \centering
  Filter \,\,($\lambda_{\mathrm eff}$)& Mag         & Error  \\
  \noalign{\smallskip}                
  \hline
  \noalign{\smallskip}                
  $g$ \,\,~\,(~0.48\,$\mu m$) & 14.729       & 0.030 \\
  $r$ \,\,~\,(~0.63\,$\mu m$) & 14.225       & 0.030 \\
  $i$ \,\,~\,(~0.77\,$\mu m$) & 14.102       & 0.030 \\
  $z$ \,\,~\,(~0.91\,$\mu m$) & 14.028       & 0.030 \\
  $J$ \,\,~\,(~1.24\,$\mu m$) & 13.142       & 0.023 \\
  $H$ \,\,\,(~1.66\,$\mu m$)  & 12.847       & 0.018 \\
  $Ks$  \,(~2.16\,$\mu m$)    & 12.799       & 0.031 \\
  $W1$  \,(~3.35\,$\mu m$)    & 12.704       & 0.025 \\
  $W2$  \,(~4.60\,$\mu m$)    & 12.771       & 0.026 \\
  $W3$  \,(11.56\,$\mu m$)    &$12.776^{a}$  &   ~~~-\\
  $W4$  \,(22.09\,$\mu m$)    &$~~9.536^{a}$ &   ~~~-\\
  \noalign{\smallskip}                
  \hline
  \end{tabular}
   \tablefoot{\tablefoottext{a}{Upper limit}}
\end{table}

Ever since June 2010 the \kepler\ team has been releasing and updating a list of transiting planet candidates, also known as \kepler\ Object of Interests (KOIs), which as of August 2014 amount to 7305 objects\footnote{Available at \url{http://exoplanetarchive.ipac.caltech.edu/cgi-bin/ExoTables/nph-exotbls?dataset=cumulative}.}. Whereas \kepler\ multi-transiting system candidates have a low probability of being false positives \citep{Lissauer2014,Rowe2014}, the same does not apply for those where a single planet is observed to transit \citep{Santerne2012,Sliski2014}. These require ground-based follow-up observations for validation, such as high-resolution spectroscopy and high-precision radial velocity measurements. The aim of follow-up  observations is thus twofold: \emph{a}) to rule out false-positive scenarios and confirm that the photometric signal is caused by a \emph{bona fide} transiting planet; \emph{b}) to characterise the system by exploiting simultaneously both the photometric and spectroscopic data.

We report on the confirmation of the \kepler\ transiting planet \object{Kepler-423b} (also known as \object{KOI-183.01}). We combined the full \kepler\ photometry with high-resolution spectroscopy from FIES@NOT to confirm the planetary nature of the transiting object and derive the system parameters. 

This paper is organised as follows. Sect.~\ref{Kepler-Photometry} describes the available \kepler\ photometry of Kepler-423, whereas Sect.~\ref{FIES-Spectroscopy} reports on our spectroscopic follow-up with FIES@NOT. In Sect.~\ref{Stellar-Properties}, we detail how the fundamental parameters of the host star were derived. In Sect.~\ref{Global-Analysis}, we outline our global Bayesian analysis and report on the quarter-to-quarter instrumental systematics affecting the \kepler\ photometry. Results are discussed in Sect.~\ref{Results} and conclusions are given in Sect.~\ref{Conclusions}.

\section{\kepler\ photometry}
\label{Kepler-Photometry}

Kepler-423 -- whose main designations, equatorial coordinates, and optical and infrared photometry are listed in Table~\ref{StarTable} -- was previously identified as a \kepler\ planet-hosting star candidate by \citet{Borucki2011} and \citet{Batalha2013} and assigned the identifier KOI-183.

The \kepler\ photometry\footnote{Available at \url{http://archive.stsci.edu/kepler}.} of Kepler-423 covers quarters 1\,--\,17 (Q$_1$\,--\,Q$_{17}$), offering four years of nearly continuous observations, from 13 May 2009 to 11 May 2013. The short cadence (\ssc; $T_\mathrm{exp}$=58.85 sec) data are available for Q$_4$\,--\,Q$_8$ and Q$_{13}$, and encompass 190 individual transits. The long cadence (\slc; $T_\mathrm{exp}$=1765.46 sec) photometry contains the \ssc transits and 311 additional \slc-only transits observed in Q$_1$\,--\,Q$_3$, Q$_9$\,--\,Q$_{12}$, and Q$_{14}$\,--\,Q$_{17}$. 

In this work we used the \kepler\ simple aperture photometry \citep[SAP;][]{Jenkins2010}, as well as the same data processed with the new version of the pre-search data conditioning (PDC) pipeline \citep{Stumpe2012}, which uses a Bayesian maximum a posteriori (MAP) approach to remove the majority of instrumental artefacts and systematic trends \citep{Smith2012}. The iterative filtering procedure by \citet{Aigrain2004} with a 5-$\sigma$ clipping algorithm was applied to both the SAP and the PDC-MAP light curves to identify and reject further outliers. We also performed a visual inspection of the \kepler\ light curves to remove photometric discontinuities across the data gaps that coincide with the quarterly rolls of the spacecraft. The point-to-point scatter estimates for the PDC-MAP \ssc and \slc light curve are 1146~ppm (1.24~mmag) and 292~ppm (0.32~mmag), respectively (Table~\ref{Parameter-Table}).

 \begin{figure}[t] 
 \begin{center}
 \resizebox{\hsize}{!}{\includegraphics[angle=0]{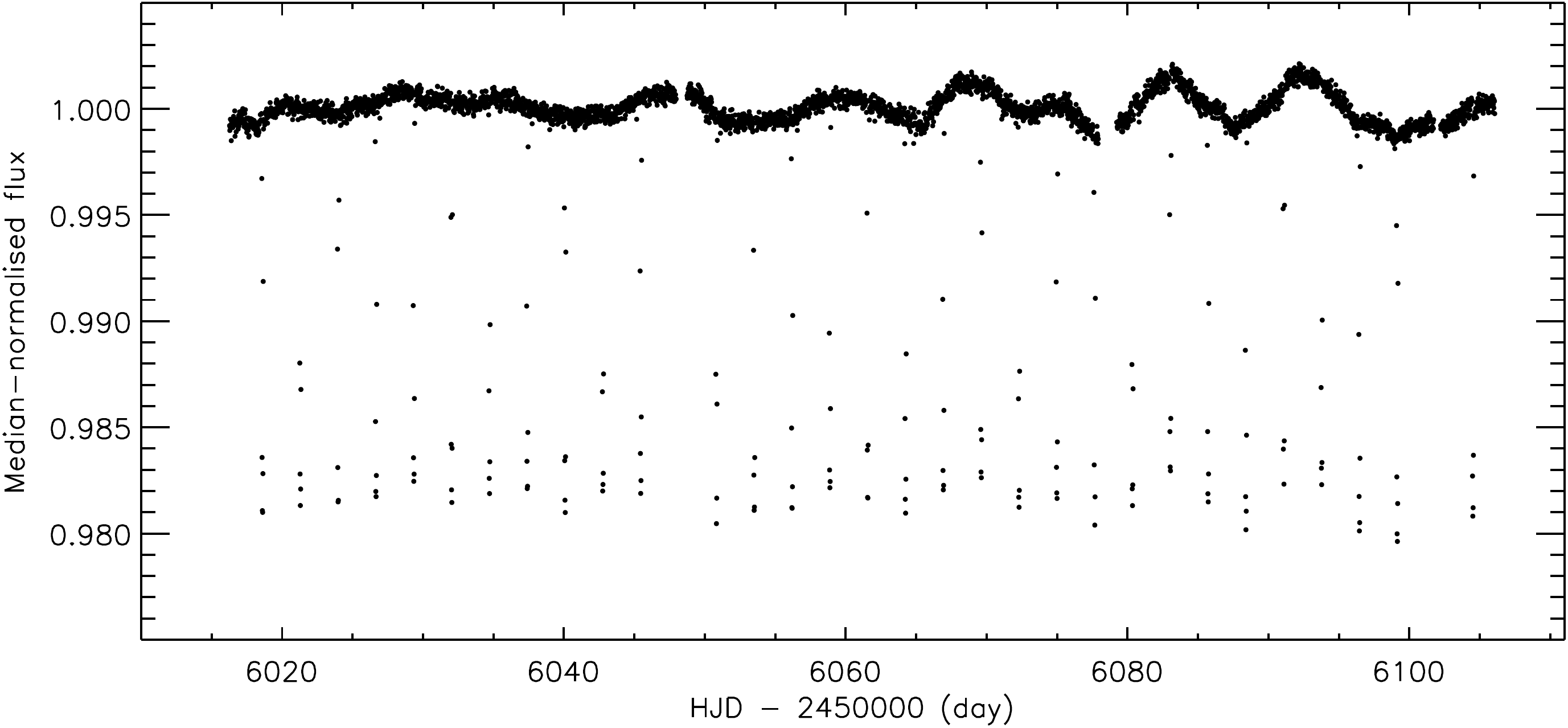}}
 \caption{Example section of median-normalised long cadence light curve of Kepler-423. Data are from \kepler\ quarter 13.}
 \label{Q13_LightCurve}
 \end{center}
 \end{figure}

Figure~\ref{Q13_LightCurve} shows the median-normalised \slc data of Kepler-423 from \kepler\ quarter 13. The $\sim$1.8\%-deep transit signals occurring every 2.7 days are clearly visible, along with a $\sim$0.5\,\% (peak-to-peak) out-of-transit modulation. Given the spectral type of the planet host star (G4\,V; see Sect.\,\ref{Photospheric-parameters}), this variability is likely to be due to magnetic active regions carried around by stellar rotation. Using an algorithm based on the autocorrelation function of the Q$_{3}$\,--\,Q$_{14}$ time-series, \citet{McQuillan2013} found a stellar rotation period of $P_\mathrm{rot}=22.047\pm0.121$~days.

\section{High-resolution spectroscopy}
\label{FIES-Spectroscopy}

\begin{table}[t]
  \centering 
  \caption{FIES radial velocity measurements of Kepler-423. The barycentric Julian dates are provided in barycentric dynamical time (BJD$_\mathrm{TDB}$). The CCF bisector spans and the S/N ratios per pixel at 5500\,\AA\ are listed in the last two columns.}
  \label{RV-Table}
\begin{tabular}{cccrc}
  \hline
  \hline
  \noalign{\smallskip}                
BJD$_\mathrm{TDB}$ &   RV    & $\sigma_{\mathrm RV}$ &   Bisector  &  S/N/pixel    \\
($-$ 2\,450\,000)  &  \kms   &    \kms               &     \kms    &  @5500\,\AA   \\
  \noalign{\smallskip}                
  \hline
  \noalign{\smallskip}                
6470.576529 & $-$3.095 & 0.040  &    0.045 & 13 \\
6472.450721 & $-$3.099 & 0.037  &    0.001 & 11 \\
6472.685581 & $-$3.141 & 0.044  &    0.049 & 10 \\
6473.557333 & $-$3.024 & 0.039  & $-$0.010 & 15 \\
6473.687627 & $-$3.024 & 0.037  & $-$0.005 & 16 \\
6485.482378 & $-$3.026 & 0.022  &    0.018 & 21 \\
6485.632534 & $-$3.039 & 0.028  &    0.026 & 18 \\
6486.453320 & $-$3.122 & 0.025  &    0.019 & 19 \\
6486.635293 & $-$3.117 & 0.028  &    0.018 & 18 \\
6487.476759 & $-$2.958 & 0.021  &    0.002 & 22 \\
6487.640452 & $-$2.927 & 0.027  & $-$0.002 & 18 \\
6544.396913 & $-$2.955 & 0.030  &    0.032 & 13 \\
  \noalign{\smallskip}                
  \hline
\end{tabular}
\end{table}

The spectroscopic follow-up of Kepler-423 was performed with the FIbre-fed \'Echelle Spectrograph \citep[FIES;][]{Frandsen1999,Telting2014} mounted at the 2.56-m Nordic Optical Telescope (NOT) of Roque de los Muchachos Observatory (La Palma, Spain). The observations were carried out between June and September 2013, under OPTICON and CAT observing programmes 2013A025 and 79-NOT14/13A, respectively. We used the $1.3\,\arcsec$ \emph{high-res} fibre, which provides, in conjunction with a 50-$\mu m$ slit at the fibre exit, a resolving power of R\,=\,67000 in the spectral range 3600\,--\,7400\,\AA. Three consecutive exposures of 1200 seconds were taken per epoch observation to remove cosmic ray hits. Following the observing strategy described in \citet{Buchhave2010}, we traced the RV drift of the instrument by acquiring long-exposed (T$_\mathrm{exp}$=15 sec) ThAr spectra right before and after each epoch observation. The data were reduced using a customised IDL software suite, which includes bias subtraction, flat fielding, order tracing and extraction, and wavelength calibration. Radial velocity measurements were derived via multi-order cross-correlation technique with the RV standard star \object{HD\,182572} -- observed with the same instrument set-up as the target object -- and for which we adopted an heliocentric radial velocity of $-100.350$\,\kms, as measured by \citet{Udry1999}.

\begin{figure}[t] 
\begin{center}
\resizebox{\hsize}{!}{\includegraphics[angle=0]{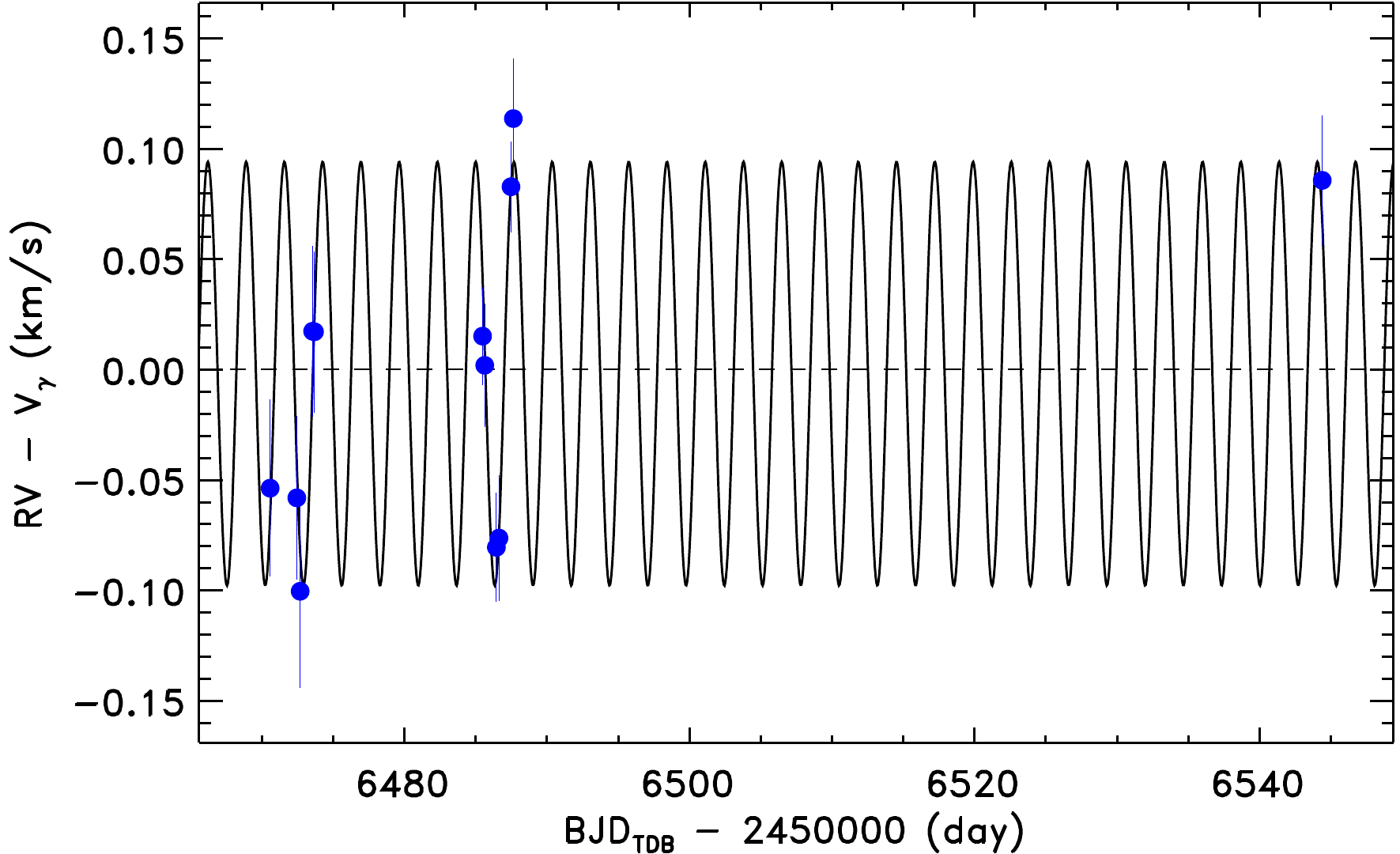}}
~\\
\resizebox{\hsize}{!}{\includegraphics[angle=0]{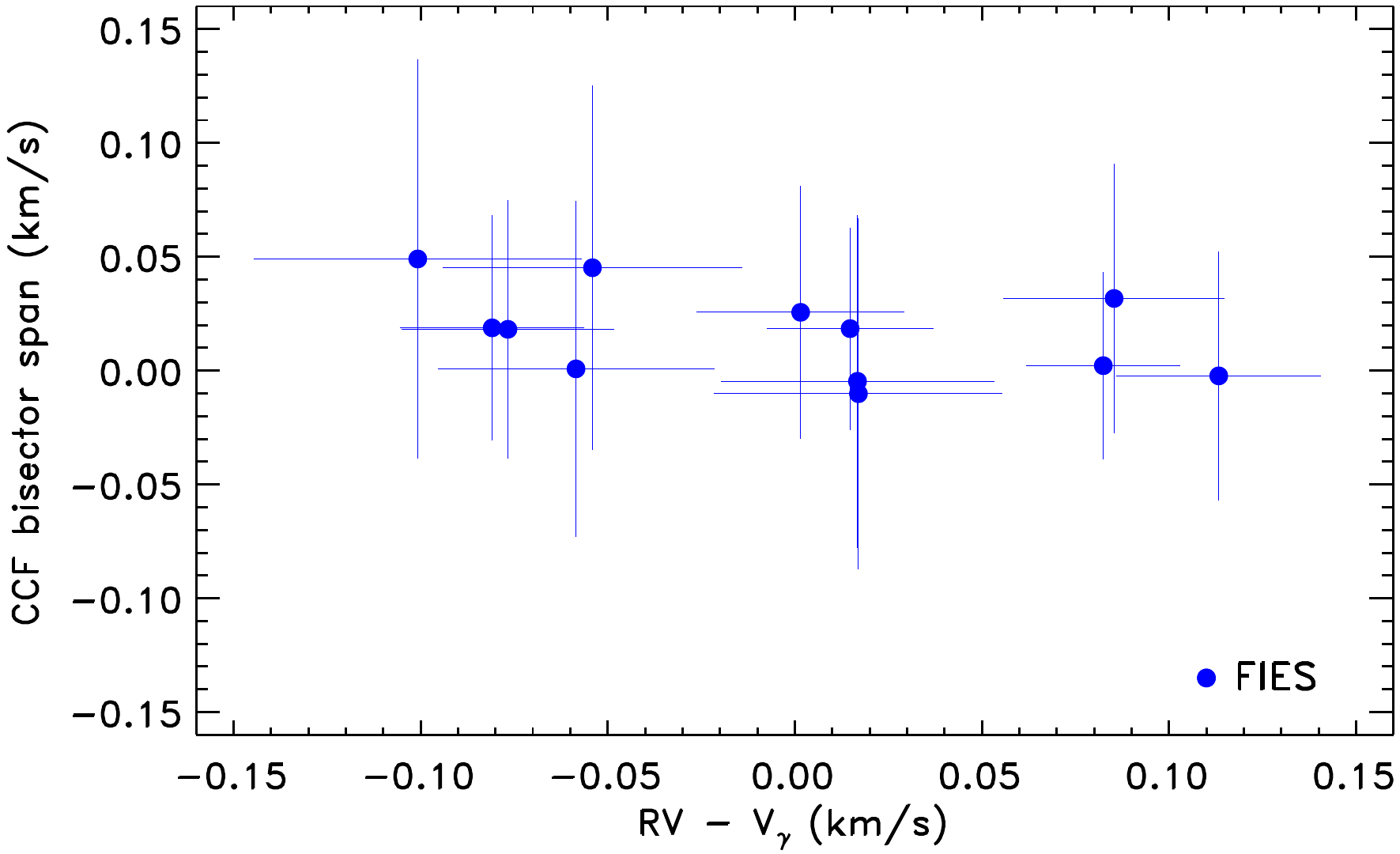}}
\caption{\emph{Upper panel}: FIES radial velocities of Kepler-423 -- after subtracting the systemic velocity $V_\gamma$ -- versus BJD$_\mathrm{TDB}$ and Keplerian fit to the data. \emph{Lower panel}: Bisector spans of the FIES cross-correlation functions versus RV measurements, after subtracting the systemic velocity $V_\gamma$. Error bars in the CCF bisector spans are taken to be twice the uncertainties in the RV~data.}
\label{RV-curve}
\end{center}
\end{figure}

The FIES RV measurements are listed in Table~\ref{RV-Table} along with the observation barycentric Julian dates in barycentric dynamical time \citep[BJD$_\mathrm{TDB}$, see][]{Eastman2010}, the cross-correlation function (CCF) bisector spans, and the signal-to-noise (S/N) ratios per pixel at 5500~\AA. The upper panel of Fig.~\ref{RV-curve} shows the FIES RVs of Kepler-423 and the Keplerian fit to the data -- as obtained from the global analysis described in Sect.~\ref{Global-Analysis} -- whereas the lower panel displays the CCF bisector spans plotted against the RV measurements, assuming that the error bars of the former are twice those of the latter. We followed the method described in \citet{Loyd2014} to account for the uncertainties of our measurements and found that the probability that uncorrelated random datasets can reproduce the observed arrangement of points (null hypothesis) is about 50\,\%. The lack of significant linear correlation between the CCF bisector spans and the RVs indicates that most likely the Doppler shifts observed in Kepler-423 are induced by the orbital motion of the companion rather than stellar activity or a blended eclipsing binary \citep[see, e.g.,][]{Queloz2001}.

\section{Properties of the host star}
\label{Stellar-Properties}

\subsection{Photospheric parameters}
\label{Photospheric-parameters}

We derived the fundamental photospheric parameters of the host star Kepler-423 using the co-added FIES spectrum, which has a S/N ratio of about 60 per pixel at 5500~\AA. Two independent analyses were performed. The first method compares the co-added FIES spectrum with a grid of theoretical models from \citet{Castelli2004}, \citet{Coelho2005}, and \citet{Gustafsson2008}, using spectral features that are sensitive to different photospheric parameters. We adopted the calibration equations for Sun-like dwarf stars from \citet{Bruntt2010} and \citet{Doyle2014} to determine the microturbulent $v_ {\mathrm{micro}}$ and macroturbulent $v_{\mathrm{macro}}$ velocities, respectively. The projected rotational velocity \vsini\ was measured by fitting the profile of several clean and unblended metal lines. The second method relies on the use of the spectral analysis package SME\,2.1, which calculates synthetic spectra of stars and fits them to observed high-resolution spectra \citep{Valenti1996,Valenti2005}. It uses a non-linear least squares algorithm to solve for the model atmosphere parameters. The two analysis provided consistent results well within the errors bars. The final adopted values are \teff~$=5560\pm80$\,K, log\,g~$=4.44\pm0.10$~(log$_{10}$\,\cms2), [M/H]~$=-0.10\pm0.05$\,dex, $v_ {\mathrm{micro}}=1.0\pm0.1$\,\kms, $v_{\mathrm{macro}}=2.8\pm0.4$\,\kms, and  \vsini\,$=2.5\pm0.5$\,\kms. Using the \citet{Straizys1981} calibration scale for dwarf stars, the effective temperature of Kepler-423 translates to a G4\,V spectral type.

\subsection{Stellar mass, radius, and age}

\begin{figure}[t] 
\begin{center}
\resizebox{\hsize}{!}{\includegraphics[angle=0]{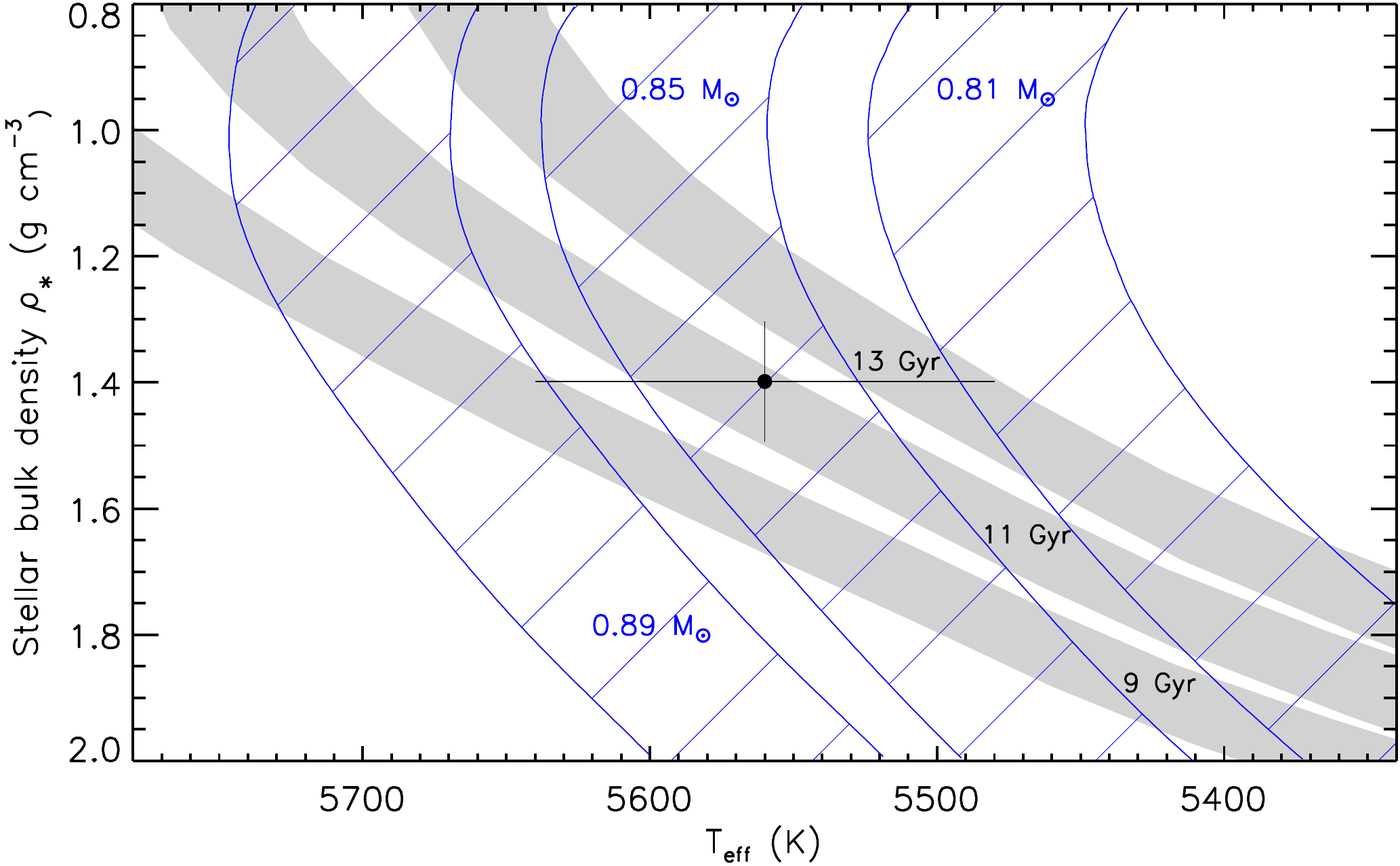}}
\caption{Modified Hertzsprung-Russell diagram showing the stellar bulk density versus effective temperature. The position of Kepler-423 is over-plotted on theoretical evolutionary tracks and isochrones from the Pisa Stellar Evolution Data Base for low-mass stars. The blue hatched areas represent different masses (0.81, 0.85, and 0.89\,\Msun\ from right to left), while the greys represent the age isochrones (9, 11, and 13 Gyr from bottom to top), computed for an initial metal content between Z=0.012 and Z=0.014.}
\label{KOI183_HR_diagram}
\end{center}
\end{figure}

Stellar mass, radius, and age were determined using the effective temperature \teff\ and metallicity [M/H], as derived from the spectral analysis (Sect.~\ref{Photospheric-parameters}), along with the stellar bulk density $\rho_\star$, as obtained from the modelling of the transit light curve (Sect.~\ref{Global-Analysis}). We compared the position of Kepler-423 on a $\rho_\star$-versus-\teff\ diagram with a grid of \emph{ad hoc} evolutionary tracks (Fig.~\ref{KOI183_HR_diagram}).

We generated stellar models using an updated version of the \emph{FRANEC} code \citep{DeglInnocenti2008,Tognelli2011} and adopting the same input physics and parameters as those used in the Pisa Stellar Evolution Data Base for low-mass stars\footnote{Available at \url{http://astro.df.unipi.it/stellar-models/}.} \citep[see, e.g.,][for a detailed description]{DellOmodarme2012}. To account for the current surface metallicity of Kepler-423 ([M/H]~$=-0.10\pm0.05$\,dex) and microscopic diffusion of heavy elements towards the centre of the star, we computed evolutionary tracks assuming an initial metal content of Z=0.010, Z=0.011, Z=0.012, Z=0.013, Z=0.014, and Z=0.015. The corresponding initial helium abundances, i.e., Y=0.268, 0.271, 0.273, 0.275, 0.277, and 0.279, were determined assuming a helium-to-metal enrichment ratio $\Delta$Y/$\Delta$Z=2 \citep{Jimenez2003,Casagrande2007,Gennaro2010} and a cosmological $^4$He abundance Y$_\mathrm{p}$=0.2485 \citep{Cyburt2004,Peimbert2007a,Peimbert2007b}. For each chemical composition, we generated a very fine grid of evolutionary tracks in the mass domain $M_\star=0.70$--$1.10$\,\Msun, with step of $\Delta M_\star=0.01$\,\Msun, leading to a total of 246 stellar tracks.

We found that evolutionary tracks with initial metal content between Z=0.012 and Z=0.014 have to be used to reproduce the current photospheric metallicity of Kepler-423. We derived a mass of $M_\star=0.85\pm0.04$\,\Msun\, a radius of $R_\star=0.95\pm0.04$\,\Rsun\, and an age of $t=11\pm2$\,Gyr (Table~\ref{Parameter-Table}). Mass and radius imply a surface gravity of log\,g~$=4.41\pm0.04$~(log$_{10}$\,\cms2), which agrees with the spectroscopically derived value log\,g~$=4.44\pm0.10$~(log$_{10}$\,\cms2).

Using pre-main sequence (PMS) evolutionary tracks would lead to consistent results in terms of stellar mass and radius, but would also yield an age of $25\pm5$~Myr. Given the relatively rapid evolutionary time-scale of PMS stars, we note that the likelihood of finding Kepler-423 still contracting towards the zero-age main sequence (ZAMS) is about 600 times lower than the probability for the star to be found in the post ZAMS phase. Moreover, such a young scenario is at odds with: \emph{a}) the distance from the galactic plane, which amounts to $166\pm17$\,pc (given the spectroscopic distance of $725\pm75$\,pc -- see below -- and galactic latitude of +12.92\,\degr); \emph{b}) the relatively long rotation period of the star ($P_\mathrm{rot}=22.047\pm0.121$~days); \emph{c}) the absence of high magnetic activity level (the peak-to-peak photometric variation is $\sim$0.5\%); \emph{d}) the lack of detectable Li\,{\sc i} $\lambda$6708~{\AA} absorption line in the co-added FIES spectrum. Short rotation period ($P_\mathrm{rot} \lesssim 5$~days), coupled with high magnetic activity and strong Li\,{\sc i} $\lambda$6708~{\AA} absorption line (EW$_\mathrm{Li} \gtrsim 300$\,m\AA), are usually regarded as youth indicators in PMS low-mass stars \citep[see, e.g.,][]{Marilli2007,Gandolfi2008}.

\subsection{Interstellar extinction and distance}

We followed the method described in \citet{Gandolfi2008} to derive the interstellar extinction $A_\mathrm{v}$ and spectroscopic distance $d$ of the system. We simultaneously fitted the available optical and infrared colours listed in Table~\ref{StarTable} with synthetic theoretical magnitudes obtained from the \emph{NextGen} model spectrum with the same photospheric parameters as the star \citep{Hauschildt99}. We excluded the $W3$ and $W4$ WISE magnitudes, owing to the poor photometry \citep{Cutri2012}. Assuming a normal extinction ($R_\mathrm{v}=3.1$) and a black body emission at the star's effective temperature and radius, we found that Kepler-423 suffers a negligible interstellar extinction of $A_\mathrm{v}=0.044\pm0.044$\,mag and that its distance is $d=725\pm75$\,pc (Table~\ref{Parameter-Table}).

\section{Bayesian and MCMC global analysis}
\label{Global-Analysis}

\begin{table}[t]
\caption{Model parametrisation used in the basic system characterisation.}
\label{table:parameterization}      
\centering  
\begin{tabular*}{\columnwidth}{@{\extracolsep{\fill}} lc}
  \hline
  \hline
  \noalign{\smallskip}                
Model Parameter                                            & Notation \\
  \noalign{\smallskip}                
  \hline
  \noalign{\smallskip}                
Planetary orbital period                                   & \pper \\
Planetary mid-transit epoch                                & \ptc  \\
Bulk stellar density                                       & \srho \\
Impact parameter                                           & \pim  \\
Orbit eccentricity                                         & \pec  \\
Argument of periastron                                     & \pom  \\
\\
Planet-to-star area ratio                                  & \paa  \\
Planet-to-star surface brightness ratio\tablefootmark{a}   & \pfr \\
\\
Linear limb-darkening coefficient                          & $u_1$ \\
Quadratic limb-darkening coefficient                       & $u_2$ \\
\kepler\ LC data scatter                                   & \pelc \\
\kepler\ SC data scatter                                   & \pesc \\
\\
Systemic radial velocity                                   & $V_\gamma$ \\
Radial velocity semi-amplitude                             & $K$ \\
\\
Quarterly crowding metric                                  & $C_i$, with $i \in [1..17]$ \\
  \noalign{\smallskip}                
  \hline
\end{tabular*}
\tablefoot{
\tablefoottext{a}{We defined the planet-to-star surface brightness ratio \pfr as the flux ratio per projected unit area (instead of as eclipse depth). The eclipse depth is thus $\Delta F_{\mathrm{ec}} = \pfr \times \paa$.}
}
\end{table}

\subsection{Approach}

\begin{figure*}
 \centering
 \includegraphics[width=\textwidth]{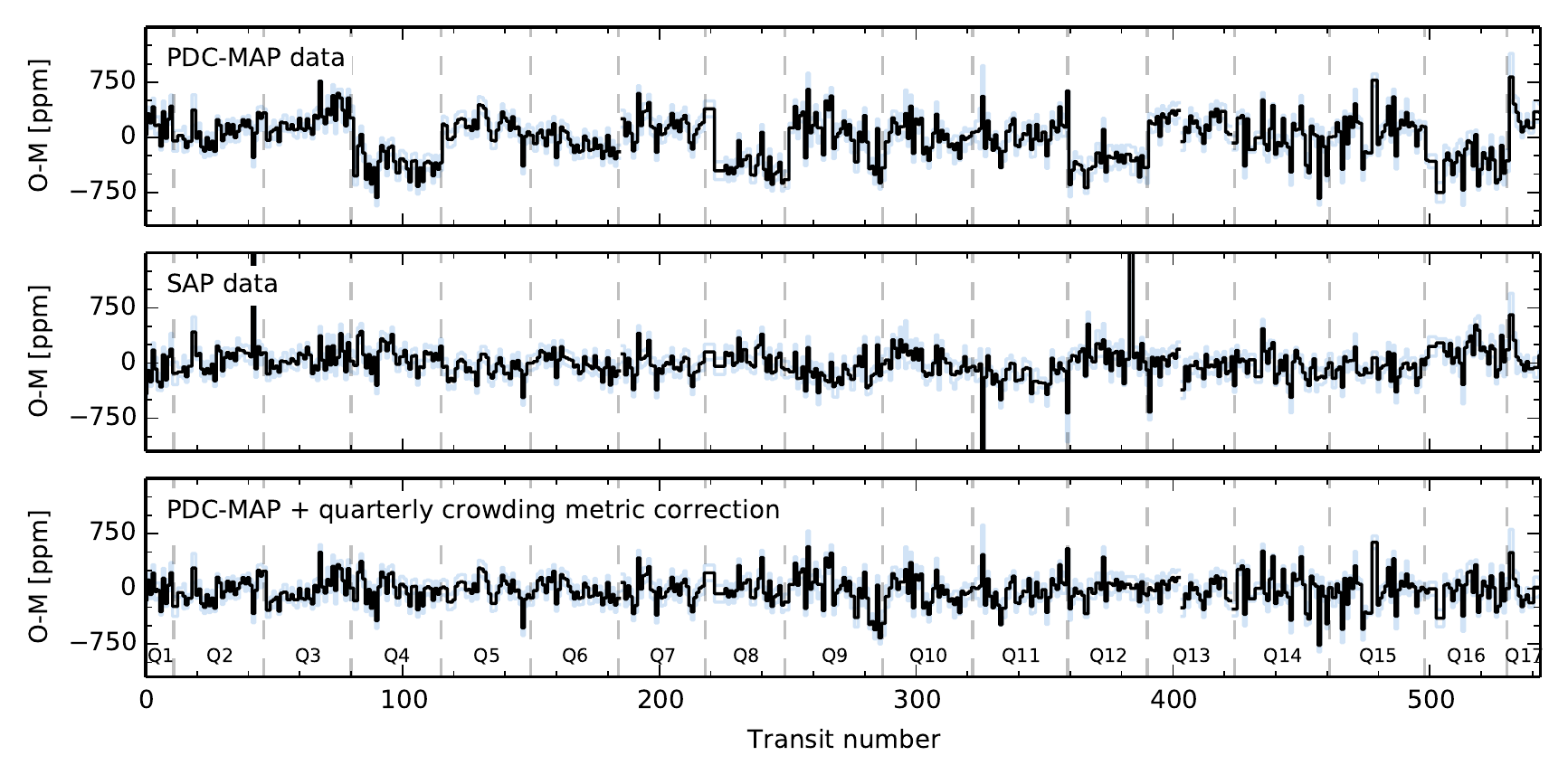}
 \caption{\emph{Upper panel.} Average, observed$-$modelled (O\,$-$\,M) flux residual of the photometric points encompassing the second ($T_2$) and third transit contact ($T_3$) as a function of the transit numbers for the PDC-MAP data, assuming that the \kepler\ contamination metric has been properly estimated. The light blue lines show the standard error of the mean for each transit. The beginning of each \kepler\ quarter is marked with dashed grey vertical lines. \emph{Middle panel.} As in the upper panel, but for the SAP data. \emph{Lower panel.} Same as before, but for the PDC-MAP cotrended data, following our quarterly crowding metric correction constrained by an informative prior (see text for more details).}
 \label{TransitDepthVariation}
\end{figure*}

We estimated the system parameters, i.e., stellar, planetary, and orbital parameters for which inference can be made based on photometry and radial velocities, using a Bayesian approach where the photometric and RV data are modelled simultaneously, similarly to the work described in \citet{Gandolfi2013} and \citet{Parviainen2014}. The model describes the primary transits, secondary eclipses, and RV variations. The significance of a possible secondary eclipse signal (Sect.~\ref{Secondary-Transit}) was assessed separately using a method based on Bayesian model comparison \citep{Parviainen2013}.

We obtained an estimate of the model posterior distribution using the Markov chain Monte Carlo (MCMC) technique. The sampling was carried out using \textit{emcee}\footnote{Available at \url{github.com/dfm/emcee}.} \citep{Foreman-Mackey2012}, a Python implementation of the Affine Invariant Markov chain Monte Carlo sampler \citep{Goodman2010}. We used \textit{PyDE}\footnote{Available at \url{github.com/hpparvi/PyDE}.}, a Python implementation of the differential evolution algorithm for global optimisation, to generate an initial population of parameter vectors clumped close to the global posterior maximum used to initialise the MCMC sampling. The sampling was carried out with 500 simultaneous walkers (chains). The sampler was first run iteratively through a burn-in period consisting of 20 runs of 500 steps each, after which the walkers had converged to sample the posterior distribution. The chains were considered to have converged to sample the posterior distribution after the ensemble properties of the chains did not change during several sets of 500 iterations, and the results from different walker subsets agreed with each other. The final sample consists of 1500 iterations with a thinning factor of 50 (chosen based on the average parameter autocorrelation lengths to ensure that we had independent samples), leading to 15000 independent posterior samples.

\subsection{Dataset}
\label{Dataset}

The dataset consists of the 12 FIES RVs (Sect.~\ref{FIES-Spectroscopy}), subsets of the \ssc and \slc data for the transit modelling, and subsets of the \slc data for the secondary eclipse modelling.

The photometric data for the transit modelling included 12 hours of data around each transit, where each segment was detrended using a second-order polynomial fitted to the out-of-transit points. We preferred short time cadence light curves when available, and excluded the \slc transits for which \ssc data was available. The final \ssc and \slc transit light curves contain about 138400 and 12100 points, respectively. We chose not to use PDC-MAP data because of the issues in the crowding metric correction applied by the pipeline, but used the PDC-MAP cotrended fluxes instead (see Sect.~\ref{KeplerSistematics}).

The eclipse model was evaluated using \slc data alone. We included about 18~hours of data centred on half-phase from each individual orbit -- enough to allow for eccentricities up to 0.2 -- and did not detrend the individual data segments (we used Gaussian processes to model the baseline instead). We rejected 69 subsets of \slc data because of clear systematics and performed the secondary eclipse modelling using 447 \slc segments. 

\subsection{Log-posterior probability density and parametrisation}
\label{ProbDensity_and_Param}

The non-normalised log-posterior probability density is described as
\begin{equation}
\begin{split}
\log P(\pv|D) = & ~~~\, \log P(\pv) \\
&+ \log P(F_\ssc|\pv) + \log P(F_\slc|\pv) \\
&+ \log P(F_\ecl|\pv) \\
&+ \log P(RV|\pv),
\label{Log-post-prob}
\end{split}
\end{equation}
where $F_\ssc$ and $F_\slc$ are the short- and long-cadence photometric data for the primary transit, $F_\ecl$ is the long cadence photometric data for the secondary eclipse, $RV$ corresponds to the FIES radial velocity data, \pv is the parameter vector containing the parameters listed in Table~\ref{table:parameterization}, and $D$ the combined dataset. The first term in the right-hand side of Eq.~\ref{Log-post-prob}, namely $\log P(\vec{\theta})$, is the logarithm of the joint prior probability, i.e., the product of individual parameter prior probabilities, and the four remaining terms are the likelihoods for the RV and light curve data.

The likelihoods for the combined RV and photometric dataset $D$ follow the basic form for a likelihood assuming independent identically distributed errors from normal distribution
\begin{equation}
\begin{split}
 \log P(D|\pv) = & -\frac{N_\mathrm{D}}{2} \log(2\pi) - N_\mathrm{D} \log(\sigma_\mathrm{D}) \\
&- \frac{1}{2} \sum_{i=1}^{N_\mathrm{D}} \left ( \frac{D_i - M(t_i,\vec{\theta})}{\sigma_\mathrm{D}} \right )^2,
\end{split}
\end{equation}
where $D_i$ is the single observed data point $i$, $M(t_i,\vec{\theta})$ the model explaining the data, $t_i$ the centre time for a data point $i$, $N_\mathrm{D}$ the number of data points, and $\sigma_\mathrm{D}$ the standard deviation of the error distribution \citep[see, e.g.,][]{Gregory2005}. 

The likelihood for the secondary eclipse data was calculated using Gaussian processes (GPs) to reduce our sensitivity to systematic noise \citep{Rasmussen2006,Gibson2012}. We modelled the residuals as a GP with an exponential kernel, with the kernel hyper-parameters fixed to values optimised to the data.

The radial velocity model follows from equation 
\begin{equation}
 RV = V_\gamma + K[\cos(\omega + \nu) + e\cos\omega],
\end{equation}
where $V_\gamma$ is the systemic velocity, $K$ the radial velocity semi-amplitude, $\omega$ the argument of periastron, $\nu$ the true anomaly, and $e$ the eccentricity. 

The transit model used PyTransit, an optimised implementation of the \citep{Gimenez2006} transit shape model\footnote{Available at \url{github.com/hpparvi/PyTransit}.}. The long-cadence and planetary eclipse models were super-sampled using 8 subsamples per \slc exposure to reduce the effects from the extended integration time \citep{Kipping2010}.

We defined the planet-to-star surface brightness ratio \pfr as the flux ratio per projected unit area (instead of as eclipse depth). The eclipse depth is thus $\Delta F_{\mathrm{ec}} = \pfr \times \paa$. 

\subsection{Systematic effects in the Kepler photometric data: quarterly transit depth variation}
\label{KeplerSistematics}

\begin{figure}[t]
 \centering
 \resizebox{\hsize}{!}{\includegraphics[angle=0]{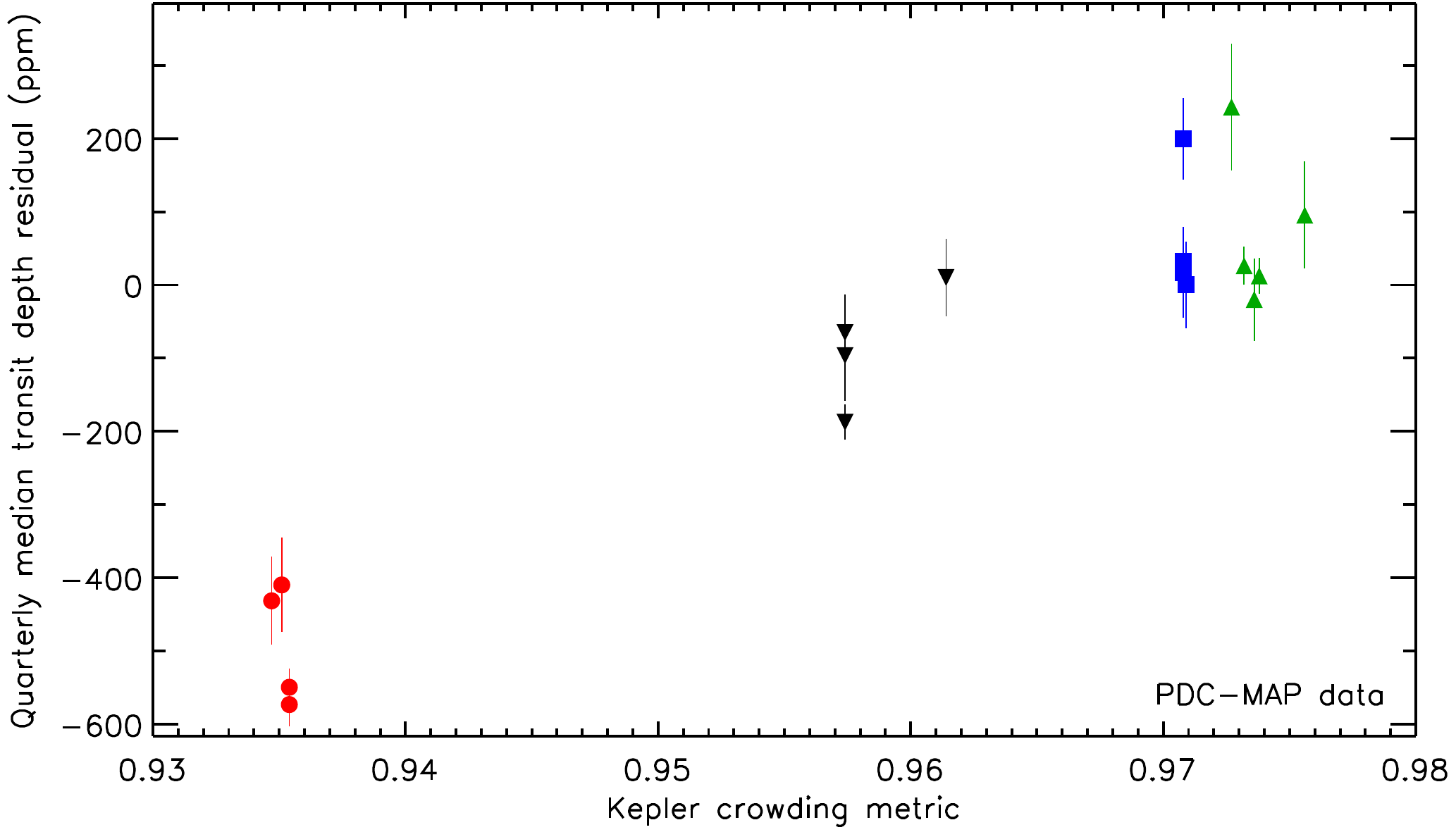}}
 ~\\
 \resizebox{\hsize}{!}{\includegraphics[angle=0]{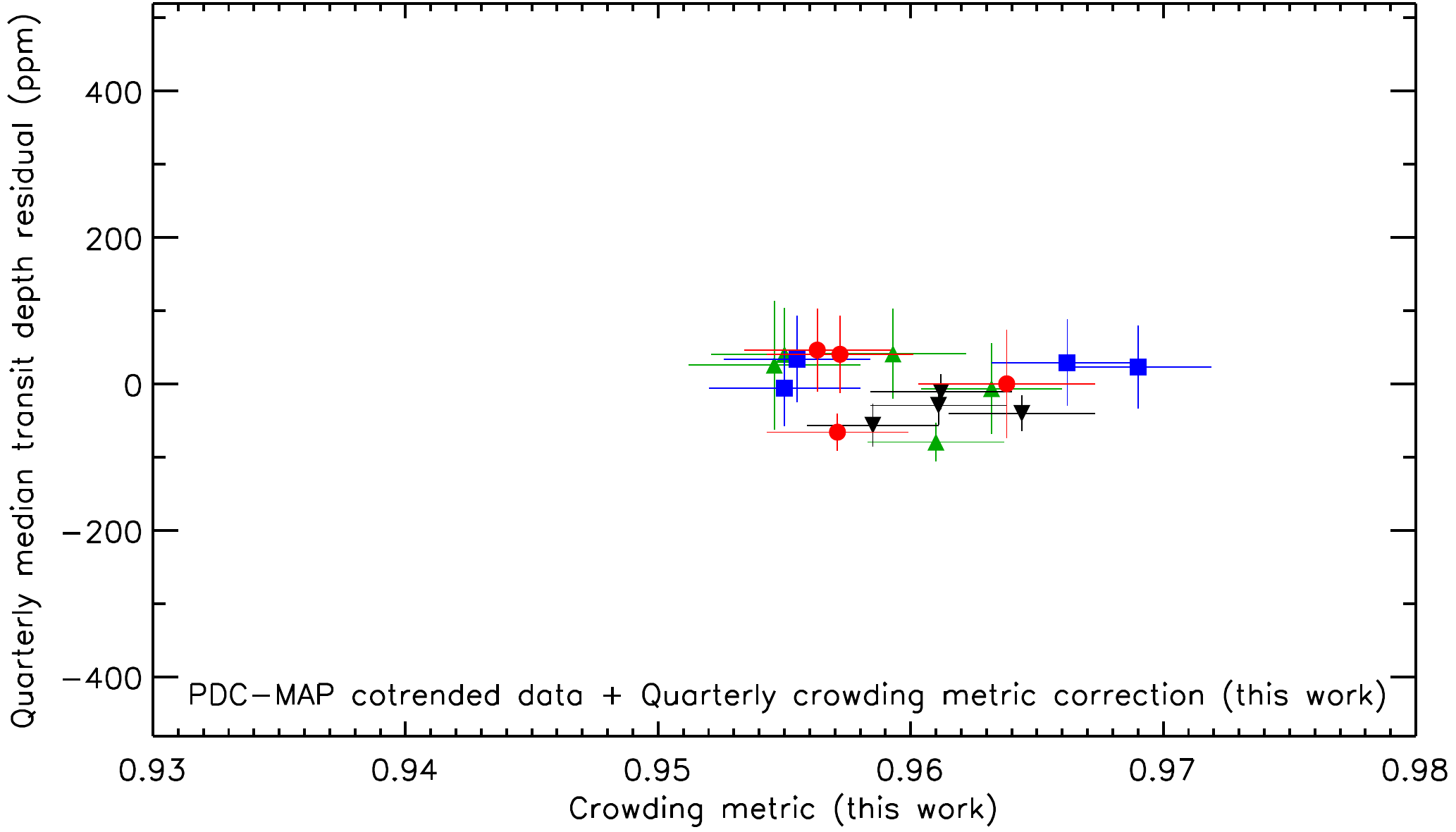}}
 \caption{\emph{Upper panel}: Quarterly median transit depth residuals, as derived from the \kepler\ PDC-MAP light curve of Kepler-423, against \kepler\ crowding metrics. Error bars are the median absolute deviations. Quarters sharing the same \kepler\ observing season are plotted with the same symbol and colour: Q$_1$, Q$_5$, Q$_9$, Q$_{13}$, and Q$_{17}$ (upward green triangles); Q$_2$, Q$_6$, Q$_{10}$, and Q$_{14}$ (downward black triangles); Q$_3$, Q$_7$, Q$_{11}$, and Q$_{15}$ (blue squares); Q$_4$, Q$_8$, Q$_{12}$, and Q$_{16}$ (red circles). \emph{Lower panel}: Same as before, but for the PDC-MAP cotrended data, following our quarterly crowding metric correction constrained by an informative prior. The x-axis reports our estimates of the quarterly crowding metrics (Table~\ref{Crowding-Metric-Table}).}
 \label{Depth_vs_CrowMetr}
\end{figure}

\citet{VanEylen2013} recently observed systematic depth variations in the \kepler\ transit light curves of \object{HAT-P-7}, which were found to be related to the 90-degree rolling of the spacecraft occurring every quarter (i.e., every $\sim$90 days). They proposed four possible causes for the variations, i.e., unaccounted-for light contamination, too small aperture photometric masks, instrumental non-linearities, and colour-dependence in the pixel response function, but noted that it is not possible to choose the most likely cause based on \kepler\ data alone.

We searched for similar instrument systematics in the \kepler\ light curve of Kepler-423 by subtracting, from each PDC-MAP and SAP transit light curve, the corresponding best-fitting transit model obtained using simultaneously all the \kepler\ segments (Sect.~\ref{Dataset}). The upper panel of Fig.~\ref{TransitDepthVariation} displays the transit depth residual as a function of the transit number for the PDC-MAP data. We found a significant ($\sim$16-$\sigma$) quarter-to-quarter systematic variation of the transit depth, with a seasonal trend reoccurring every four quarters and with the $Q_{4}$, $Q_{8}$, $Q_{12}$, and $Q_{16}$ data yielding the deepest transit light curves. The peak-to-peak amplitude is about 800 parts per million (ppm), which corresponds to $\sim$4.3\,\% of the mean transit depth. 

Intriguingly, there is no significant ($\sim$2-$\sigma$) quarter-to-quarter variation of the transit depth in the SAP data, as shown in the middle panel of Fig.~\ref{TransitDepthVariation}. The SAP residuals exhibit, however, intra-quarter systematic trends that might result from the motion of the target within its photometric aperture due to telescope focus variation, differential velocity aberration, and spacecraft pointing \citep{Kinemuchi2012}. The PDC-MAP data are corrected for these effects using cotrending basic vectors generated from a suitable ensemble of highly-correlated light curves on the same channel \citep{Stumpe2012}. Because different behaviours of the \kepler\ detectors would most likely cause systematics visible in both PDC-MAP and SAP data, we can safely exclude the channel-to-channel non-linearity difference as the source of the quarter-to-quarter transit depth variation. Moreover, the \kepler\ CCD non-linearity is reported to be 3\,\% over the whole dynamic range \citep{Caldwell2010} and the systematic variations are at least one order of magnitude larger than the expected non-linearity effect at the transit depth signal.

\begin{table}[t]
\centering 
\caption{\kepler\ quarterly crowding metrics (second column). Our estimates of the crowding metrics with their 1-$\sigma$ uncertainties from the MCMC posterior sampling are listed in the last two columns.}
\begin{tabular}{cccc}
  \hline
  \hline
  \noalign{\smallskip}                
Quarter  & \kepler\ crowding & Derived crowding  & $\sigma_{\mathrm C_i}$ \\
 (Q$_i$) &      metric       &   metric ($C_i$)  &                        \\
  \noalign{\smallskip}                
  \hline
  \noalign{\smallskip}                
  Q$_{1}$ & 0.9756 & 0.9638  &  0.0035 \\
  Q$_{2}$ & 0.9614 & 0.9572  &  0.0029 \\
  Q$_{3}$ & 0.9708 & 0.9563  &  0.0029 \\
  Q$_{4}$ & 0.9354 & 0.9571  &  0.0028 \\
  Q$_{5}$ & 0.9738 & 0.9644  &  0.0029 \\
  Q$_{6}$ & 0.9574 & 0.9612  &  0.0028 \\
  Q$_{7}$ & 0.9708 & 0.9611  &  0.0027 \\
  Q$_{8}$ & 0.9354 & 0.9585  &  0.0026 \\
  Q$_{9}$ & 0.9736 & 0.9690  &  0.0029 \\
 Q$_{10}$ & 0.9574 & 0.9550  &  0.0030 \\
 Q$_{11}$ & 0.9709 & 0.9662  &  0.0030 \\
 Q$_{12}$ & 0.9347 & 0.9555  &  0.0029 \\
 Q$_{13}$ & 0.9732 & 0.9610  &  0.0027 \\
 Q$_{14}$ & 0.9574 & 0.9593  &  0.0029 \\
 Q$_{15}$ & 0.9708 & 0.9632  &  0.0028 \\
 Q$_{16}$ & 0.9351 & 0.9550  &  0.0029 \\
 Q$_{17}$ & 0.9727 & 0.9546  &  0.0034 \\
  \noalign{\smallskip}                
  \hline  
\end{tabular}
\label{Crowding-Metric-Table}
\end{table}

This leaves the crowding metric correction performed by the PDC-MAP pipeline as the most plausible explanation. The crowding metric is defined as the fraction of light in the photometric aperture arising from the target star. Since apertures are defined for each quarter -- to account for the redistribution of target flux over a new CCD occurring at each roll of the spacecraft -- the crowding metrics are computed quarterly for each target. The excess flux due to crowding within the photometric aperture is automatically removed by the PDC-MAP pipeline from the SAP light curve. The upper panel of Figure\,\ref{Depth_vs_CrowMetr} shows the quarterly median transit depth residuals -- as derived from the PDC-MAP light curve -- plotted against the crowding metrics -- as extracted from the header keyword CROWDSAP listed in the \kepler\ data (Table~\ref{Crowding-Metric-Table}, second column). We found a significant correlation between the two quantities, with a null hypothesis probability of only 0.15\,\%. The lack of significant quarter-to-quarter transit depth variation in the SAP data (Fig.~\ref{TransitDepthVariation}, middle panel) suggests that the crowding metric variation among quarters is most likely overestimated, i.e., the excess flux arising from nearby contaminant sources is very likely to be almost constant from quarter to quarter.

\citet{Kinemuchi2012} quoted the completeness of the \kepler\ input catalogue (KIC) as possible source of contamination error. However, from a comparison with the POSS\,II image centred around Kepler-423, we noted that all nearby faint stars up to \kepler\ magnitude $K_p < 20$~mag are included in the KIC. We therefore considered the \kepler\ crowding metric values to be generally correct, but not their variations among quarters. The crowding metric correction is the last data processing step performed by the PDC-MAP pipeline on the cotrended \kepler\ light curve \citep{Smith2012, Stumpe2012}. To account for the quarterly systematics, we carried out the parameter estimation with the PDC-MAP cotrended data. The latter were obtained by removing the \kepler\ crowding metric correction from the pipeline-generated PDC-MAP data. The parameter estimation was then performed including a per-quarter contamination metric to the model with informative prior based on the average crowding metric derived by the \kepler\ team (Fig.~\ref{TransitDepthVariation}, lower panel). Thus, our approach also yielded fitted estimates for the quarterly crowding factor. 

\subsection{Priors}
\label{Priors}

\begin{figure}[t]
 \centering
 \includegraphics[width=\columnwidth]{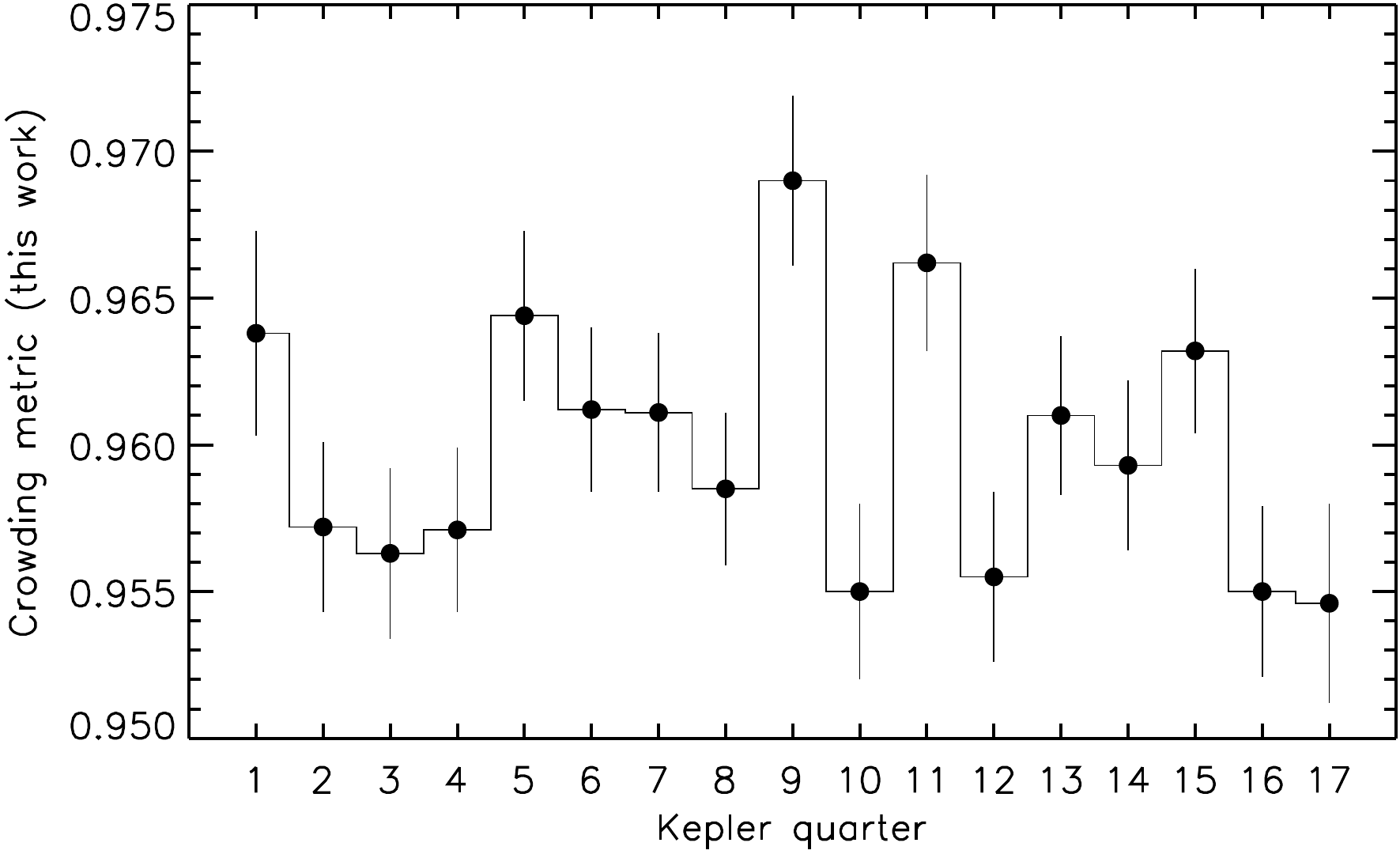}
 \caption{Quarterly crowding metric estimates rederived from our analysis and their 1-$\sigma$ uncertainties (Table~\ref{Crowding-Metric-Table}) versus quarter numbers. The posterior crowding estimates are a product of transit modelling using the PDC-MAP cotrended data and informative priors based on the \kepler\ crowding estimates.}
 \label{Crowding-Metric-Plot}
\end{figure}

The final joint model has 31 free parameters, listed in Table~\ref{table:parameterization}. We used uninformative priors (uniform) on all parameters except the 17 crowding metrics, for which we used normal priors centred on 0.96 with a standard deviation of 0.01, based on the crowding metrics estimated by the \kepler\ team. While reducing the objectivity of the analysis, setting an informative prior on the quarterly crowding metrics was a necessary compromise, since the shape of the transit light curve alone cannot constrain totally free contamination.

\begin{figure*}[t]
 \centering
 \includegraphics[width=\textwidth]{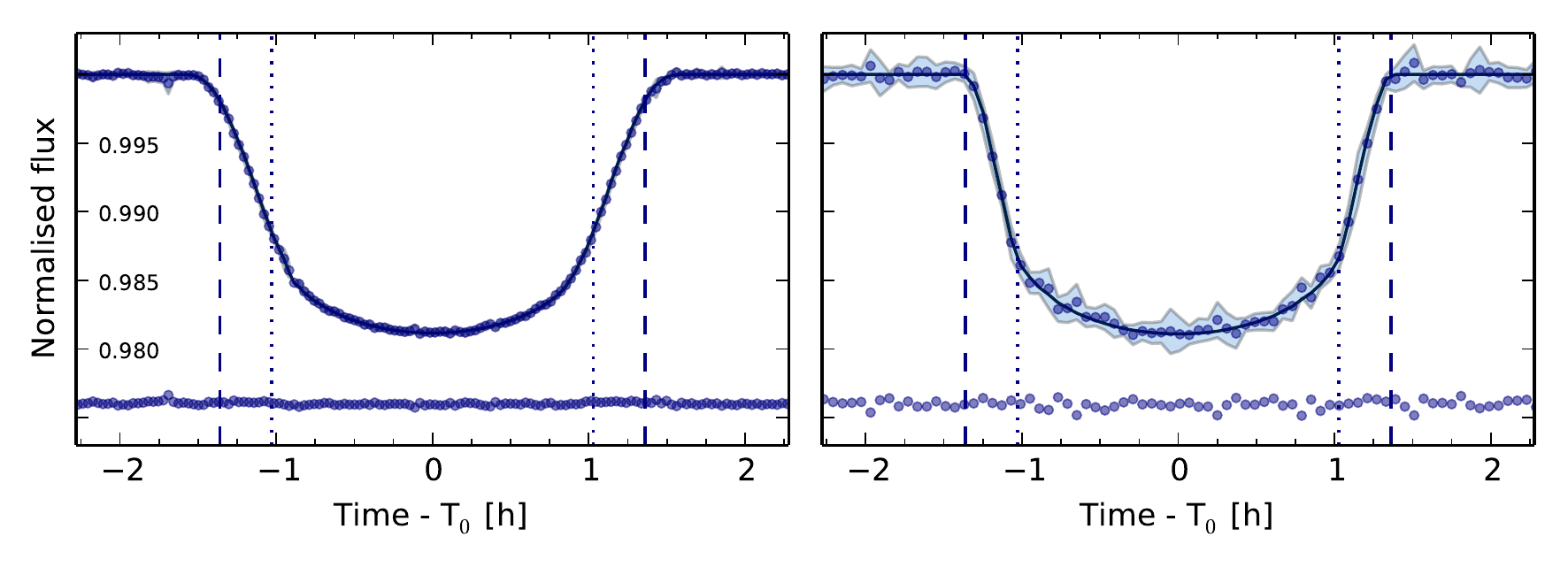}
 \caption{Phase-folded transit light curves of Kepler-423, best fitting model, and residuals. \slc data are shown on the left panel, \ssc on the right panel, both binned at $\sim$1.9 minutes. The shaded area corresponds to the 3-$\sigma$ errors in the binned fluxes. The dashed lines mark the $T_{14}$ limits, and the dotted lines the $T_{23}$ limits. The blurring of the transit shape -- due to the long integration time -- is obvious in the \slc plot (left panel).}
 \label{Transit-Lightcurve}
\end{figure*}

We considered two cases for the secondary eclipse. For model comparison purposes, we carried out the sampling for a model with a delta prior forcing the planet-to-star surface brightness ratio to zero (no-eclipse model), and with a uniform prior based on simple modelling of expected flux ratios. We estimated the allowed range for the planet-to-star surface brightness ratio using a Monte Carlo approach by calculating the flux ratios for 50000 samples of stellar effective temperature, semi-major axis, heat redistribution factor, and Bond albedo. The effective temperature and semi-major axis distributions are based on estimated values (Table~\ref{Parameter-Table}), the heat distribution factor values are drawn from uniform distribution $U(1/4,2/3)$ and the Bond albedo values from uniform distribution $U(0,0.5)$. The resulting distribution is nearly uniform, and extends from 0.0 to 0.012 (99$^\mathrm{th}$ percentile), and thus we decided to set a uniform prior $U(0,0.012)$ on the surface brightness ratio.

\section{Results and discussions}
\label{Results}

\begin{figure}[t]
 \centering
 \includegraphics[width=\columnwidth]{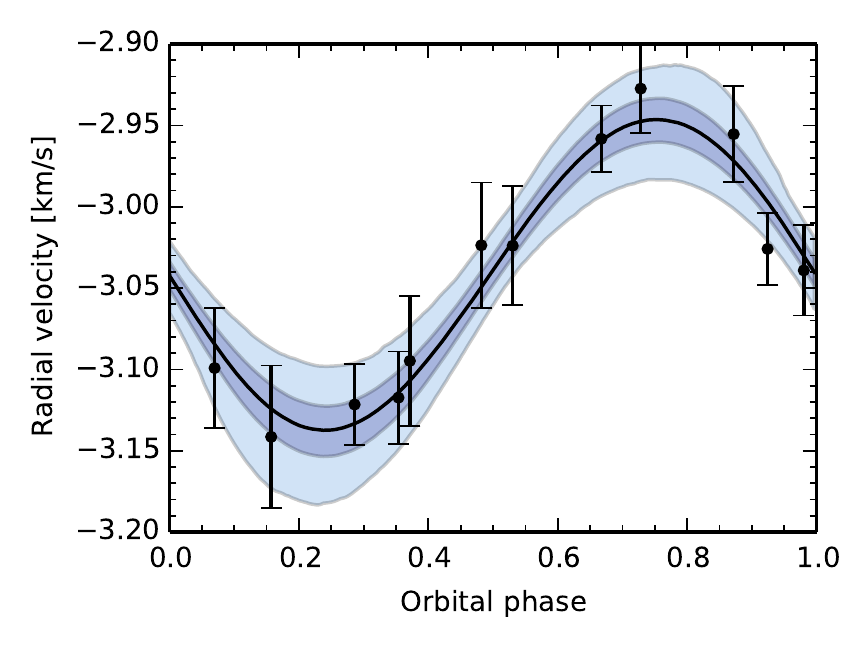}
 \caption{Radial velocity data with the median and 68\% and 99\% percentile limits of the posterior predictive distribution.}
 \label{Phase-Folded-RV-Curve}
\end{figure}

We list our results in Table~\ref{Parameter-Table}. The system's parameter estimates were taken to be the median values of the posterior probability distributions. Error bars were defined at the 68\,\% confidence limit. We show the phase-folded transit and RV curves along with the fitted models in Figs.~\ref{Transit-Lightcurve} and \ref{Phase-Folded-RV-Curve}, respectively.

The quarterly correction estimates, along with their 1-$\sigma$ uncertainties, are listed in Table~\ref{Crowding-Metric-Table}. For the sake of illustration, they are also plotted in Fig.~\ref{Crowding-Metric-Plot}. We found no significant correlation (51\,\% null hypothesis probability) between the quarterly transit depth -- as derived from the PDC-MAP cotrended light curve, following our correction for contamination factor -- and the crowding metric, as displayed in the lower panel of Figure~\ref{Depth_vs_CrowMetr}. We note that neglecting the quarter-to-quarter transit depth variation leads to a significant (7-$\sigma$) underestimate of the planet-to-star radius ratio by about 1.5\,\% and doubles its uncertainty.

\subsection{Planet properties}
\label{Planet-properties}

\begin{figure}[t]
 \centering
 \includegraphics[width=\columnwidth]{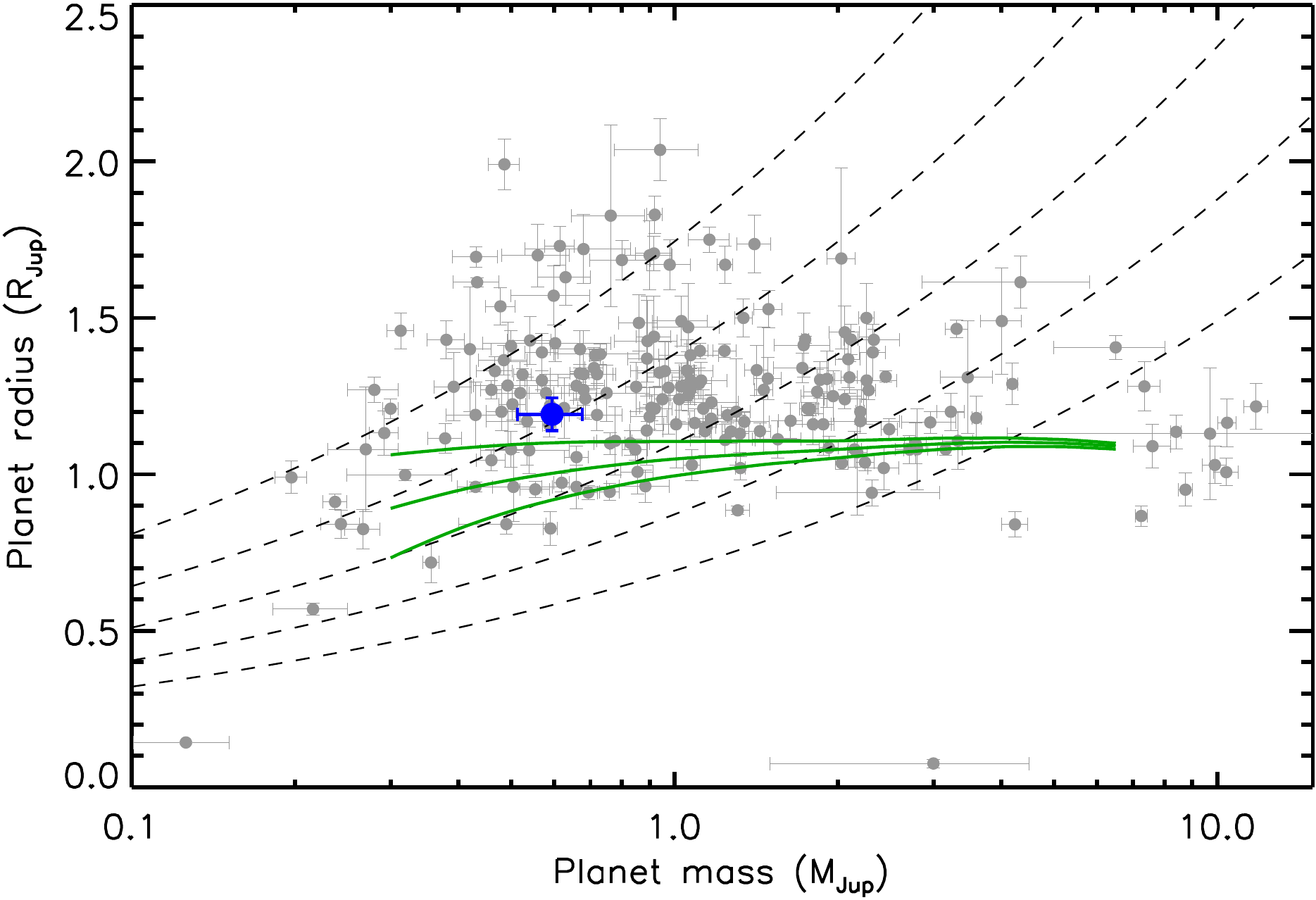}
 \caption{Mass-radius diagram for transiting hot Jupiters (grey circles; \pper\,$<10$~days and $0.1<M_\mathrm{p}<15$~\Mjup, from
the Extrasolar Planet Encyclopedia at \url{http://exoplanet.eu/}, as of 15 September 2014). Kepler-423b is marked with a thicker blue circle. The \citet{Fortney2007} isochrones for a planet core mass of 0, 25, 50~\mearth\ -- interpolated to the solar equivalent semi-major axis of Kepler-423b and extrapolated to an age of 11\,Gyr -- are overplotted with thick green lines from top to bottom. Isodensity lines for density $\rho_\mathrm{p} = 0.25$, 0.5, 1, 2, and 4~\gcm3 are overlaid with dashed lines from left to right.} 
 \label{Mass-Radius-Diagram}
\end{figure}

The planet Kepler-423b has a mass of \mp~$=0.595\pm0.081$~\Mjup\ and a radius of \rp~$=1.192\pm0.052$~\Rjup, yielding a planetary bulk density of $\rho_\mathrm{p}=0.459\pm0.083$~\gcm3. We show in Fig.~\ref{Mass-Radius-Diagram} how Kepler-423b compares on a mass-radius diagram to all other known transiting hot Jupiters (\pper\,$<10$\,days; $0.1<M_\mathrm{p}<15$~\Mjup). With a system age of 11\,Gyr, the radius of Kepler-423b is consistent within 1.5-$\sigma$ with the expected theoretical value for an irradiated coreless gas-giant planet \citep{Fortney2007}. Alternatively, the planet might have a core and be inflated because of unaccounted-for heating source, atmospheric enhanced opacities, and reduced interior heat transport \citep{Guillot2008,Baraffe2014}. It is worth noting that the radius of Kepler-423b agrees within 1-$\sigma$ to the empirical radius relationship for Jupiter-mass planets from \citet{Enoch2012}, which predicts a radius of $1.28\pm0.14$\,\Rjup, given the planetary mass $M_\mathrm{p}$, equilibrium temperature $T_\mathrm{eq}$, and semi-major axis $a_\mathrm{p}$ listed in Table~\ref{Parameter-Table}.

\subsection{Secondary eclipse and planet albedo}
\label{Secondary-Transit}

\begin{figure}[t]
 \centering
 \resizebox{\hsize}{!}{\includegraphics[angle=0]{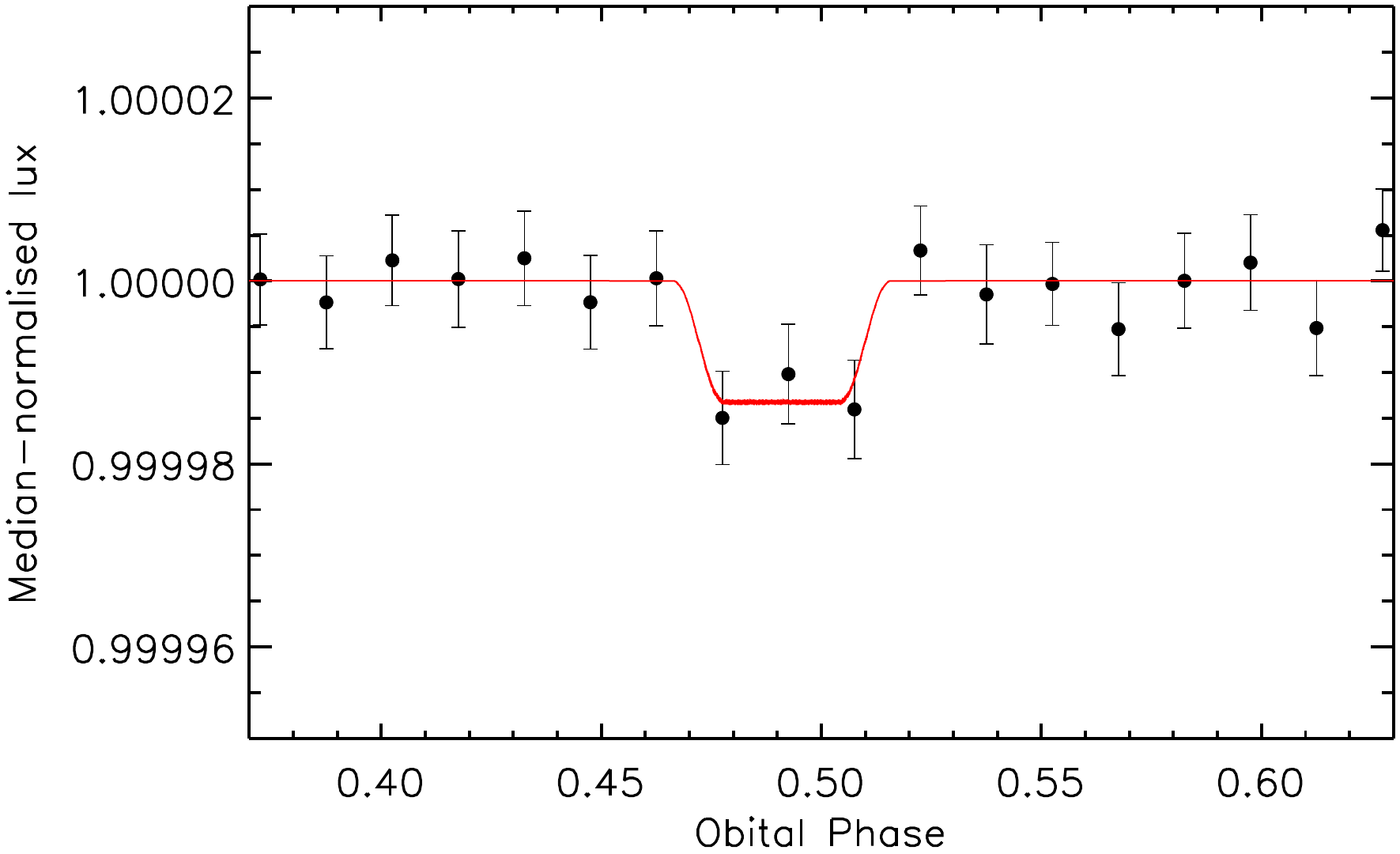}}
 \caption{Secondary eclipse light curve of Kepler-423, phase-folded to the orbital period of the planet. \kepler\ data are median-binned in intervals of 0.015 cycles in phase ($\sim$1 hour). The 1-$\sigma$ error bars are the median absolute deviations of the data points inside the bin, divided by the square root of the number of points. The best fitting transit model is overplotted with a red line.}
 \label{Secondary-Lightcurve}
\end{figure}

We detected a tentative secondary eclipse of Kepler-423b in the \kepler\ long cadence light curve and measured a depth of $14.2\pm6.6$~ppm (Fig.~\ref{Secondary-Lightcurve}). The eclipse signal is relatively weak and could in theory be due to random instrumental or astrophysical events. We set to verify the eclipse signal and assessed its significance using a method based on Bayesian model selection, as described by \citet{Parviainen2013}. We introduced some additional improvements that we briefly describe in the following paragraph\footnote{For a detailed explanation, see Parviainen et al. 2015, in preparation.}. The Bayesian evidence integration could be carried out using simple Monte Carlo integration because of the low dimensionality of the effective parameter space.

We considered two models, one without an eclipse signal ($M_0$) and one with an eclipse signal ($M_1$), and assigned equal prior weights on both models. We calculated the Bayes factor in favour of $M_1$ cumulatively for each orbit, i.e., we calculated the Bayesian evidence for both models separately for every individual 18 hour-long data segment. Since we assumed that the model global likelihoods -- or Bayesian evidence -- are independent from orbit to orbit, the final global likelihood is the product of the model likelihoods for each orbit -- or a sum of the model log-likelihoods. A real eclipse signal that exists from orbit to orbit leads to a steadily increasing Bayes factor in favour of $M_1$.\footnote{We stress that this is a slight simplification, as the cumulative Bayes factor behaves more as a directed random walk, especially for weak signals.} In contrast, a signal from an individual event mimicking an eclipse would be visible as a jump in the cumulative Bayes factor trace.

The Bayes factor in favour of the eclipse model was found to depend  strongly on our choice of priors on eccentricity and surface brightness ratio. Assuming a uniform prior on eccentricity between 0 and 0.2 and a Jeffreys prior on surface brightness ratio encompassing all physically plausible values for planetary albedos up to 0.5 (i.e., flux ratios between 0 and 0.008) results in a Bayes factor only slightly higher than unity. Lowering the maximum eccentricity to 0.05 and maximum surface brightness ratio to 0.0015 (based on our MCMC posterior sampling, which is going to the grey area of Bayesian model selection) yielded a Bayes factor of $\sim$2.6, corresponding to positive support for the eclipse model.

\begin{figure}[t]
 \centering
 \includegraphics[width=\columnwidth]{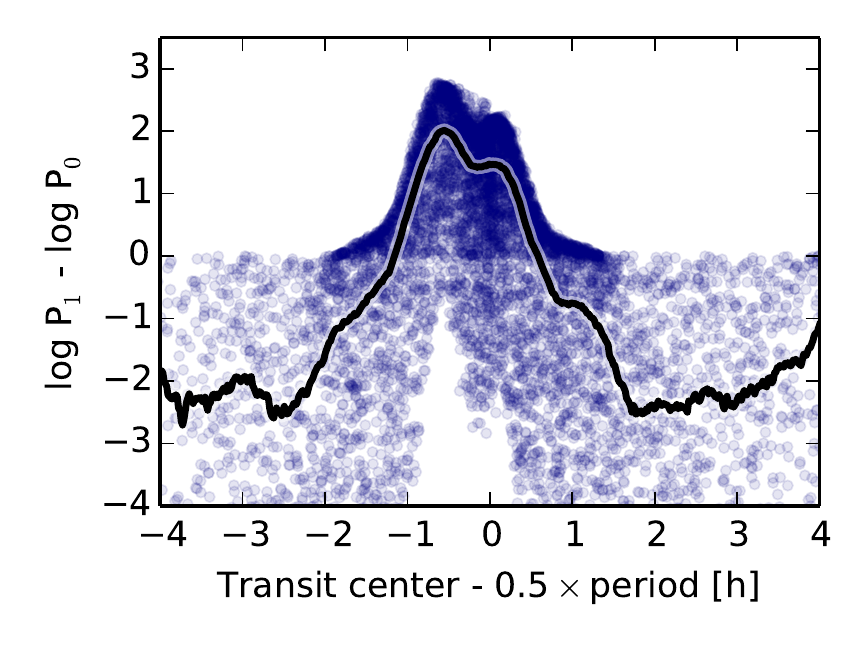}
 \caption{Differences between the individual posterior samples for the eclipse model (M$_1$) and no-eclipse model (M$_0$) plotted against the eclipse centre (light blue circles), mapped from the sampling space that uses eccentricity and argument of periastron. Only 447 \slc segments are used for the modelling (see Sect.~\ref{Dataset}). A Bayes factor map produced by sliding a uniform prior with a width of 15 minutes along the transit centre is overlaid with a black thick line.}
 \label{fig:secondary_1d}
\end{figure}

\begin{figure}[t]
 \centering
 \includegraphics[width=\columnwidth]{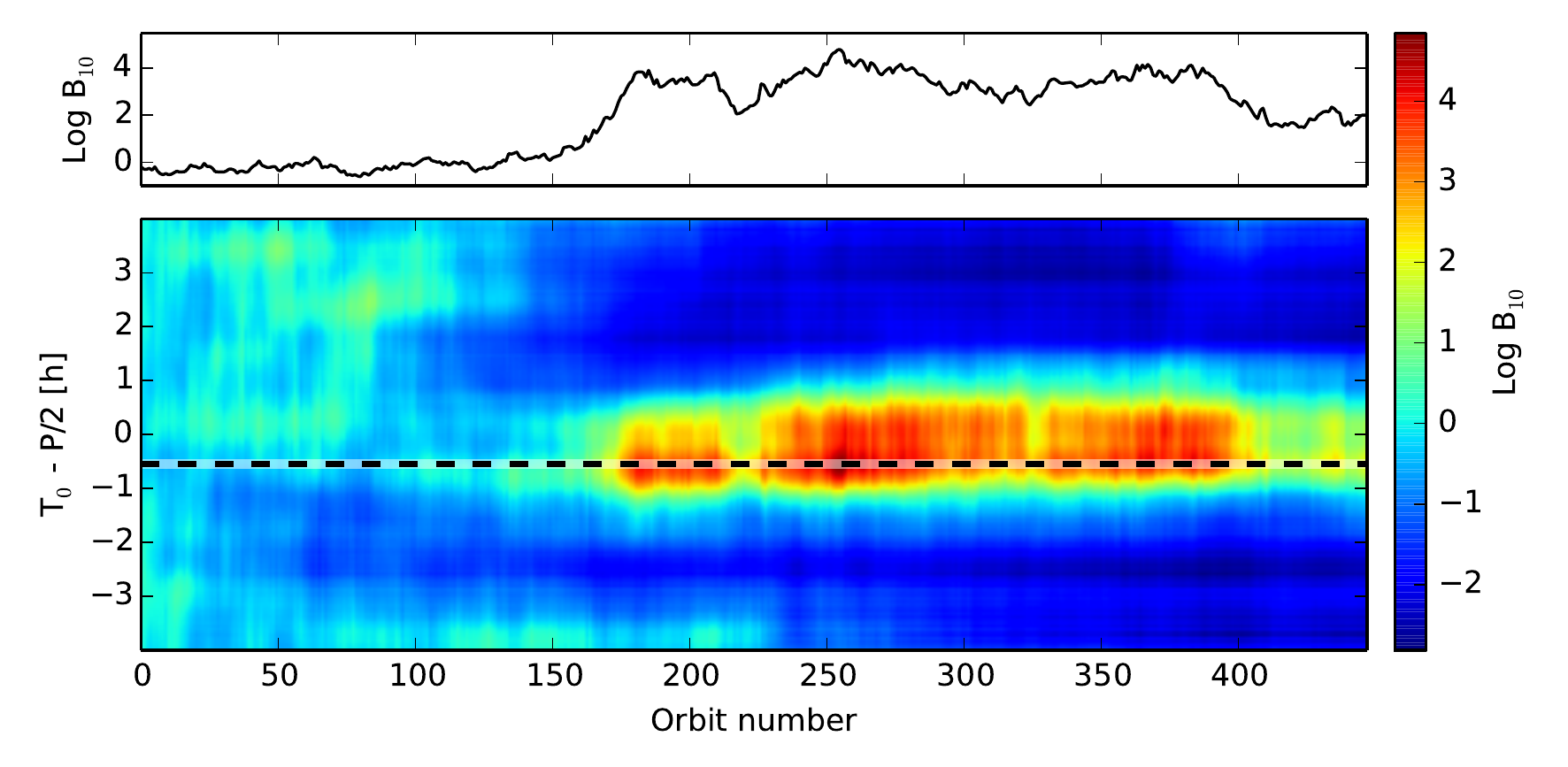}
 \caption{The cumulation of log~B$_{10}$ as a function of increasing data. The plot shows only the 447 orbits used for the secondary eclipse modelling (see Sect.~\ref{Dataset}). The upper plot shows the trace for the maximum log~B$_{10}$ case (identified as a slashed line in the lower plot), and the lower plot shows log~B$_{10}$ mapped as a function of a sliding prior on the eclipse centre (as in Fig.~\ref{fig:secondary_1d}) on the y-axis, with the amount of data (number of orbits included) increasing on the x-axis.}
 \label{fig:sec_cumbf}
\end{figure}

We show the log posteriors sample differences and the Bayes factor in favour of the eclipse model mapped as a function of eclipse centre -- itself a function of the eccentricity and argument of periastron -- in Fig.~\ref{fig:secondary_1d}, and the cumulative Bayes factor in Fig.~\ref{fig:sec_cumbf}.

The Bayes factor map is used as an expository tool to probe the Bayes-factor space as a function of our prior assumptions, and in this case shows that \emph{a}) the tentative eclipse found near 0.5 phase is the only eclipse-like signal inside the sampling volume constrained by our priors; \emph{b}) while the Bayes factor is only moderately in favour of the eclipse model, it is against the eclipse model for eclipse signals occurring away from the identified eclipse (with a peak-to-peak log Bayes factor difference being $\sim$4). However, the Bayes factor trace (Fig.~\ref{fig:sec_cumbf}) shows that the support for the eclipse-model is mostly from a small continuous subset of orbits (but not from a single orbit that would indicate a jump in the data). Thus, we must consider the detected eclipse signal to be only tentative.

The depth of the planetary eclipse would imply a planet-to-star surface brightness ratio of $\pfr=(8.93\pm4.13) \times 10^{-4}$, allowing us to constrain the geometric $A_\mathrm{g}$ and Bond $A_\mathrm{b}$ albedo of the planet. From the effective stellar temperature, eccentricity, and scaled semi-major listed in Table~\ref{Parameter-Table}, and assuming $A_\mathrm{g}=1.5 \times A_\mathrm{B}$ and heat redistribution factor between 1/4 and 2/3, we found tentative values of $A_\mathrm{g}=0.055\pm0.028$, $A_\mathrm{B}=0.037\pm0.019$, and a planet brightness temperature of $T_\mathrm{br}=1950\pm250$~K. This would make Kepler-423b one of the gas-giant planets with lowest Bond albedo known so far \citep[see, e.g.,][]{Angerhausen2014}.

\subsection{Tidal interaction and non-zero eccentricity}
\label{Tital_interaction}

The RV data alone constrain the eccentricity to $\pec<0.16$ (99$^\mathrm{th}$ percentile of the posterior distribution). The inclusion of the photometric data and the tentative detection of the planet occultation give a small non-zero eccentricity of $0.019^{+0.028}_{-0.014}$. We stress that ignoring the planet eclipse signal and imposing a circular orbit have a negligible effect on the values of the derived planetary parameters. 

We estimated the tidal evolution time-scales of the system using the model of \citet{Leconte2010}, which is valid for arbitrary eccentricity and obliquity. However, instead of using a constant time lag $\Delta t$ between the tidal bulge and the tidal potential, we recast their model equations using a constant modified tidal quality factor $Q^{\prime}$, for an easy comparison with results for other planetary systems usually given in terms of $Q^{\prime}$. Specifically, we assumed that $\Delta t = 3/(2k_{2} n Q^{\prime})$, where $k_{2}$ is the potential Love number of the second degree and $n$  the mean orbital motion. This approximation is the same as, e.g.,  in \citet{Mardling2002} and is justified for a first estimate of the time-scales in view of our limited knowledge of tidal dissipation efficiency inside stars and planets. 

The rotation period of the star is longer than the orbital period of the planet, therefore tides act to reduce the semi-major axis of the orbit $a$ and to spin up the star. Assuming $Q^{\prime}_{*} = 10^{6}$ for the star, we obtained a tidal decay time-scale $|(1/a)(da/dt)|^{-1}\sim4$~Gyr, while the time-scales for spin alignment and spin up are both  $|(1/\Omega) (d\Omega/dt)|^{-1}\sim10$~Gyr, all comparable to the age of the system. This suggests that a substantial orbital decay accompanied by a spin up of the star could have occurred during the main-sequence evolution of the system. Applying standard gyrochronology to the star \citep{Barnes2007}, we estimated an age of $\sim$3.7~Gyr for the observed rotation period of 22.046~days, which supports the conclusion that stellar magnetic braking is counteracted by the planet. The expected rotation period is $\sim$40~days for an age of 11~Gyr, i.e., the star is rotating about 1.8 times faster than expected. The present orbital angular momentum is about 2.5 times the stellar spin angular momentum. If the angular momentum of excess rotation comes from the initial orbital angular momentum, its minimum value was $\sim$1.25 the present orbital angular momentum, corresponding to an initial orbital period of at least 4.9~days for a planet of constant mass. If the stellar tidal quality factor $Q^{\prime}_{*} \ga 10^{7}$, as suggested by \citet{OgilvieLin2007} for non-synchronous systems as in the case of Kepler-423, the orbital decay and the stellar spin up would have been negligible during the main-sequence evolution of the host and the excess rotation could be associated with a reduced efficiency of the stellar wind due to the magnetic perturbations induced by the planet \citep{Lanza2010,Cohenetal2010}. 

Separately, if the orbital eccentricity is indeed non-zero and equal to its most likely value $0.019^{+0.028}_{-0.014}$, we estimated that this would require $Q^{\prime}_{*} \geq 10^{7}$ and $Q^{\prime}_{\rm p} \geq 10^{8}$. The latter is consistent with the tidal quality factor estimated by \citet{GoodmanLackner2009} for coreless planets.

\subsection{Spin-orbit alignment along the line of sight}

The spin-orbit angle, i.e., the angle between the stellar spin axis and the angular momentum vector of the orbit, is regarded as a key parameter to study planet migration mechanisms \citep[see, e.g.,][]{Winn2010b,Gandolfi2012,Crida2014}. Assuming that a star rotates as a rigid body, one can infer the inclination $i_\star$ of the stellar spin with respect to the line of sight through
\begin{equation}
 \sini =  P_\mathrm{rot}\,(\vsini) / 2\,\pi\,R_{\star},
\end{equation}
where $P_\mathrm{rot}$, $R_{\star}$, and \vsini\ are the stellar rotation period, radius, and projected rotational velocity, respectively. Since a transiting planet is seen nearly edge-on ($i_p\approx 90$\degr), the inclination of the stellar spin axis can tell us whether the system is aligned along the line of sight or not. However, the method does not allow us to distinguish between prograde and retrograde systems, as $i_\star$ and $\pi - i_\star$ angles provide both the same \sini.

Using the values reported in Table~\ref{Parameter-Table}, we found that $\sini=1.15\pm0.23$, which implies that $i_\star$ is between $\sim$70 and 90\degr\ or $\sim$160 and 180\degr. Given the fact that the measurements of the Rossiter-McLaughlin effect have shown that retrograde systems around relatively cool stars (\teff$\lesssim$6250\,K) are rare \citep[see, e.g.,][]{Winn2010b,Albrecht2012,Hirano2014}, our findings are consistent with spin-orbit alignment along the line of sight. Moreover, as seen in Sect\,~\ref{Tital_interaction}, the tidal interaction time-scale for the evolution of the obliquity is comparable to the age of the system, implying that any primordial misalignment of the planet has most likely been damped down by tidal forces. This agrees with the general trend observed in systems with short tidal interaction time-scales \citep{Albrecht2012}.

\subsection{Search for transit timing variations}
\label{TTVs}

We carried out a search for additional perturbing objects in the system by looking for gravitationally-induced variations in the transit centre times of Kepler-423b, the so-called transit timing variations (TTVs). The TTV search was carried out using MCMC and exploiting -- for the first time -- the full \kepler\ light curve of Kepler-423, from Q$_1$ to Q$_{17}$. The transit centre posteriors were estimated by fitting a transit model to the individual transits with parameter posteriors from the main characterisation run used as priors for all the parameters except the transit centre. A wide uniform prior centred on the expected transit centre time, assuming no TTVs, was used for the transit centre.

\begin{figure}[t]
 \resizebox{\hsize}{!}{\includegraphics[angle=0]{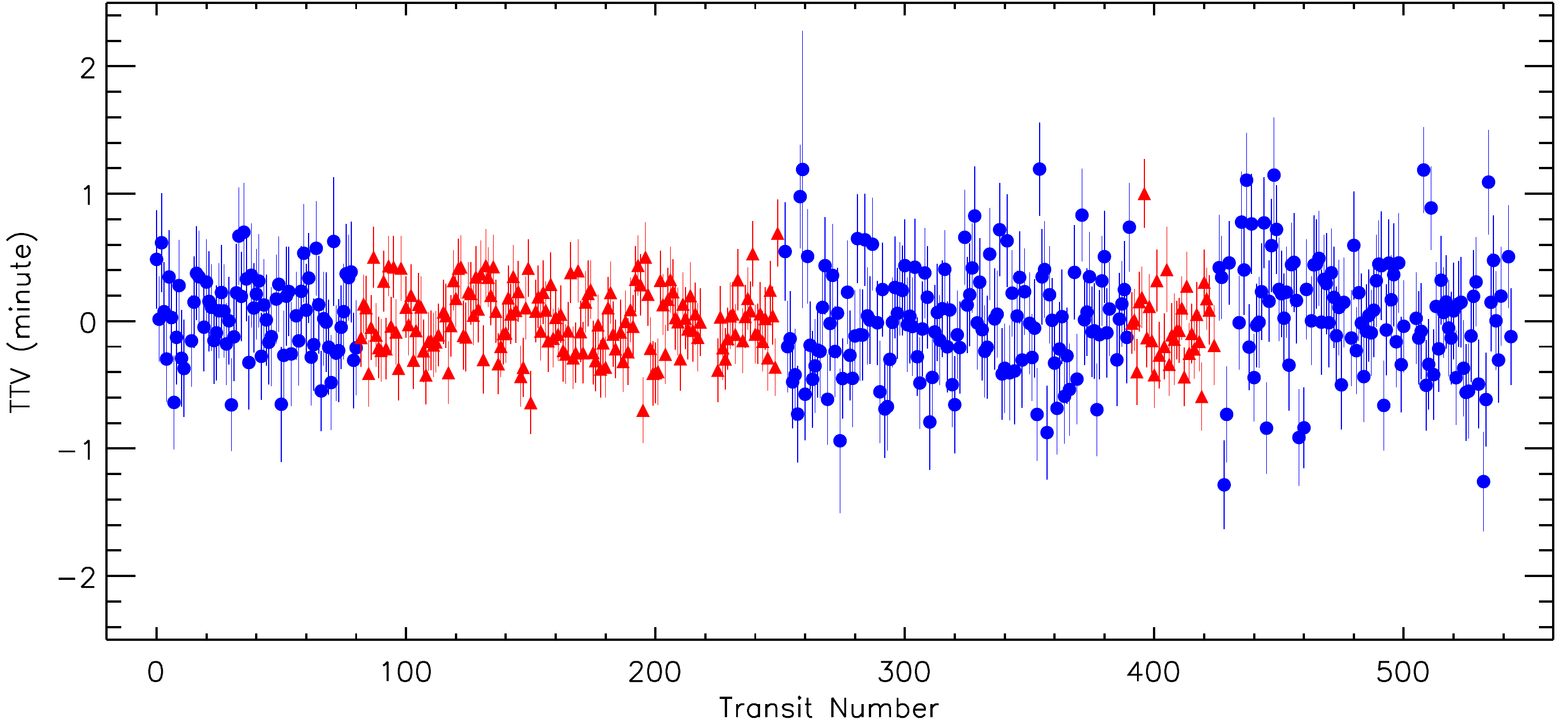}}
 ~\\
 \resizebox{\hsize}{!}{\includegraphics[angle=0]{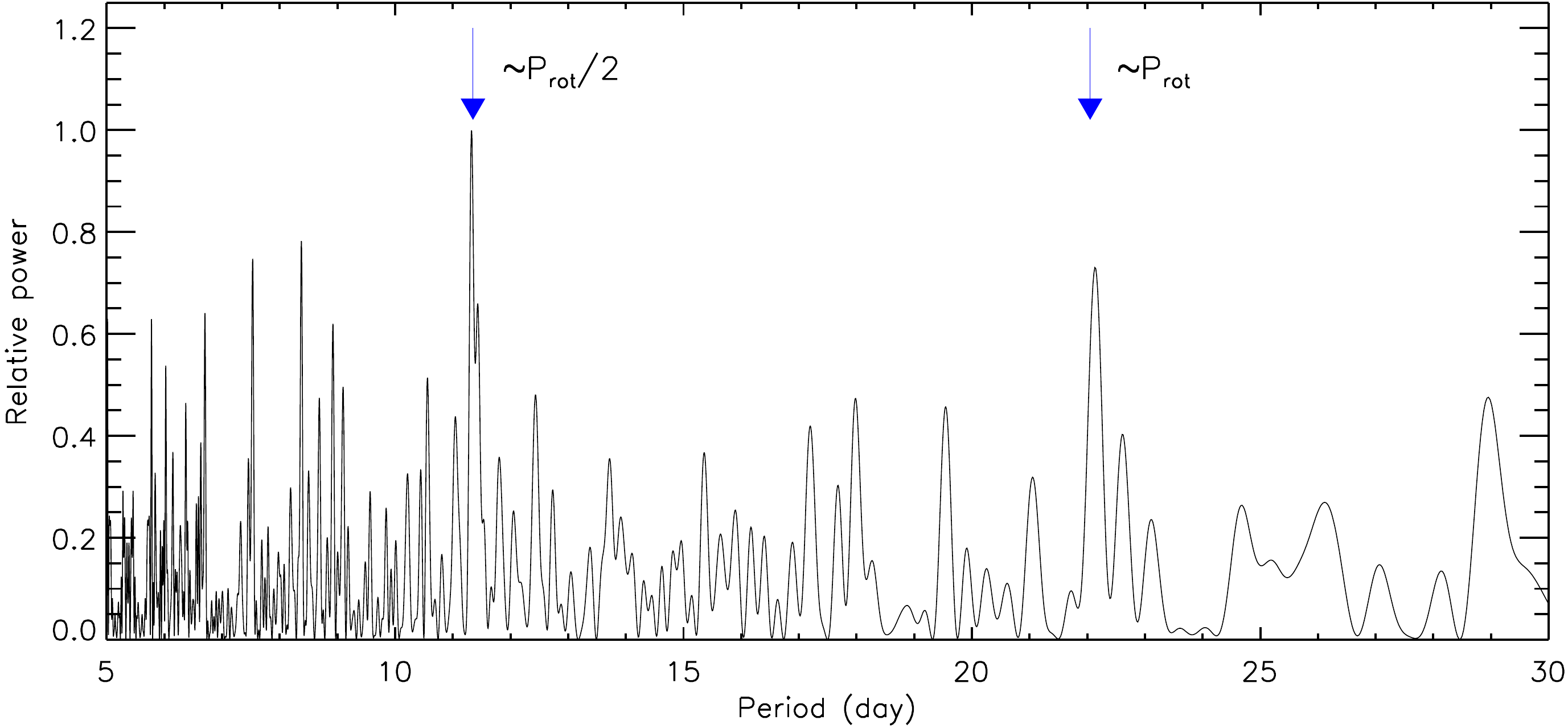}}
 ~\\
 \resizebox{\hsize}{!}{\includegraphics[angle=0]{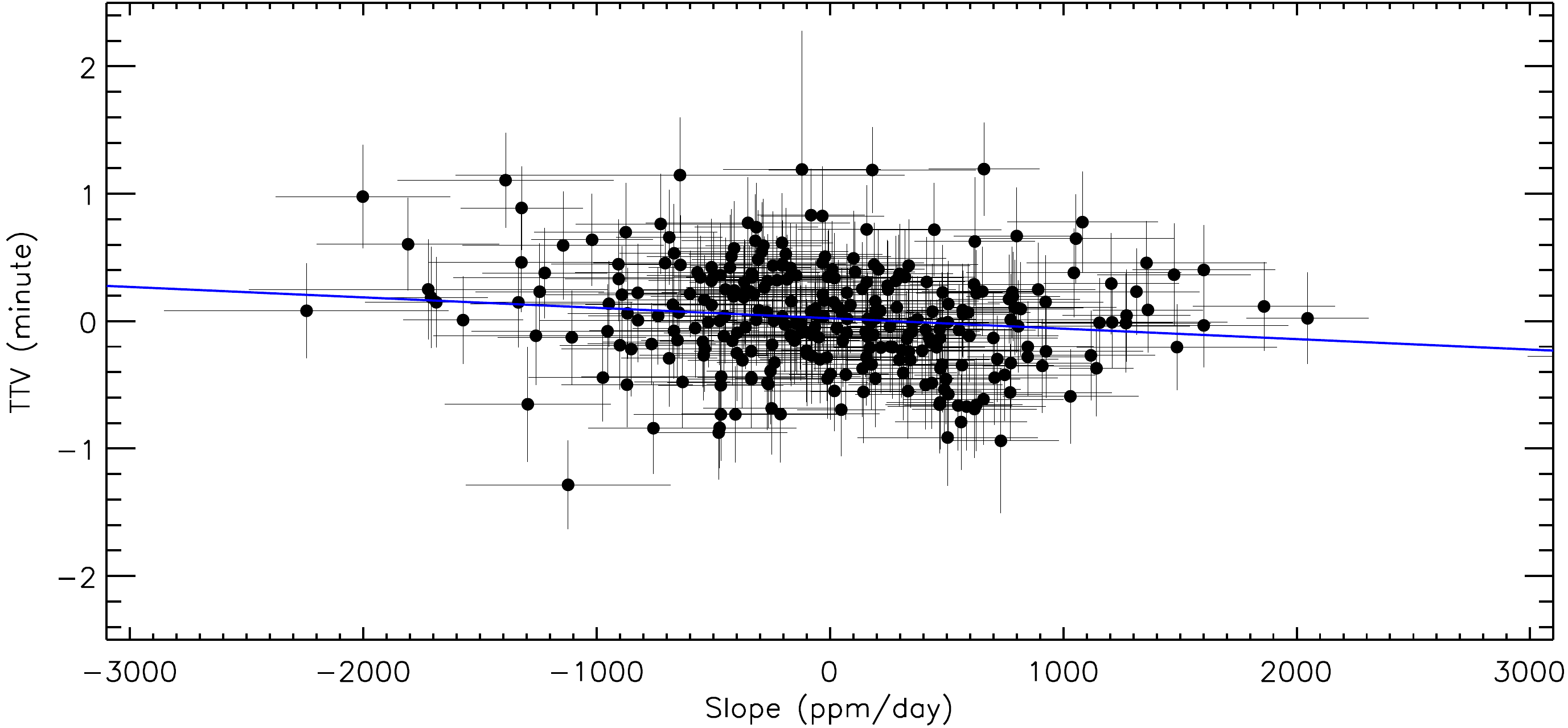}}
 \caption{\emph{Upper panel}: Differences between the observed and modelled transit centre times of Kepler-423 (TTVs). Transit timing variations extracted from the short cadence data are plotted with red triangles, while long cadence TTVs are shown with blue circles. \emph{Middle panel}: Lomb-Scargle periodogram of the TTV data in the 5--30 day period range. The two arrows mark the peak close to the stellar rotation period and its first harmonic. \emph{Lower panel}: Transit timing variation versus local transit slope. The straight blue line marks the linear fit to the data.}
 \label{KOI183-TTV}
\end{figure}

Our results are shown in the upper panel of Fig.~\ref{KOI183-TTV}. The transit centres do not deviate significantly from the linear ephemeris and the results allow us to rule out TTVs with peak-to-peak amplitude larger than about 2 minutes. Our finding agrees with those from \citet{Ford2011} and \citet{Mazeh2013}, and further confirm the trend that stars hosting hot Jupiters are often observed to have no other close-in planets \citep{Steffen2012,Szabo2013}. 

The Lomb-Scargle periodogram of the Kepler-423b TTV data shows no peaks with false alarm probability smaller than 1\,\%. However, it is worth noting that there are peaks at the stellar rotation period and its first harmonic (Fig.~\ref{KOI183-TTV}, middle panel), which might be due to the passage of Kepler-423b in front of active photospheric regions \citep[see, e.g.,][]{Oshagh2013}. The peak at half the stellar rotation period might be caused by the occultations of starspots at opposite stellar longitudes. As a matter of fact, the \kepler\ light curve shows also quasi-periodic variations recurring every $\sim$11\,days (i.e., half the stellar rotation period), which are visible in the second half of the Q$_{13}$ data plotted in Fig.~\ref{Q13_LightCurve}.

As recently suggested by \citet{Mazeh2014}, the anti-correlation (correlation) between the TTV and the slope of the light curve around each transit can be used to identify prograde (retrograde) planetary motion with respect to the stellar rotation. The lower panel of Figure~\ref{KOI183-TTV} shows that there is no significant correlation (anti-correlation) in our data between TTVs and local photometric slopes, the linear Pearson correlation coefficient and null hypothesis probability being $-$0.16 and 13\,\%, respectively.

The lack of TTV-versus-slope correlation (anti-correlation) might imply that there are no detectable spot-crossing events in the TTV data, and that the photometric variation observed in the \kepler\ light curve is mainly dominated by active regions that are not occulted by the planet. Alternatively, there might be spot-crossing events whose signal in the TTV data is just below the noise. As a sanity check, we performed a visual inspection of the SC transits -- the only ones in which a spot-crossing event can potentially be identified by eye -- and found only one significant event. However, we believe that our data are mainly dominated by noise, because the TTVs are normally distributed around zero with a standard deviation ($\sim$22~sec) comparable with the average uncertainty of our measurements ($\sim$19~sec).

\section{Conclusions}
\label{Conclusions}

We spectroscopically confirmed the planetary nature of the \kepler\ transiting candidate Kepler-423b. We derived the systems parameters exploiting -- for the first time -- the whole available \kepler\ photometry and combined it with high-precision RV measurements taken with FIES@NOT. 

We found that the PDC-MAP \kepler\ data are affected by seasonal systematic transit depth variations recurring every 4 quarters. We believe that these systematics are caused by an uncorrected estimate of the quarterly variation of the crowding metric, rather than different behaviours of the \kepler\ detectors (linearity), and treated them as such.

Kepler-423b is a moderately inflated hot Jupiter with a mass of \mp~$=0.595\pm0.081$~\Mjup\ and a radius of \rp~$=1.192\pm0.052$~\Rjup, translating into a bulk density of $\rho_\mathrm{p}=0.459\pm0.083$~\gcm3. The radius is consistent with both theoretical models for irradiated coreless giant planets and expectations based on empirical laws. Kepler-423b transits every 2.7~days an old, G4\,V star with an age of $11\pm2$\,Gyr. 

The stellar rotation period, projected equatorial rotational velocity \vsini\, and star radius $R_{\star}$ constrain the inclination $i_\star$ of the stellar spin axis to likely lie between $\sim$70 and 90\degr, implying that the system is aligned along the line of sight.

We found no detectable TTVs at a level of $\sim$22~sec (1-$\sigma$ confidence level), confirming the lonely trend observed in hot Jupiter data. Our tentative detection of the planetary eclipse yields a small non-zero eccentricity of $0.019^{+0.028}_{-0.014}$, and geometric and Bond albedo of $A_\mathrm{g}=0.055\pm0.028$ and $A_\mathrm{b}=0.035\pm0.014$, respectively, placing Kepler-423b amongst the gas-giant planets with the lowest albedo known so far.

\begin{acknowledgements}
We are infinitely grateful to the staff members at the Nordic Optical Telescope for their valuable and unique support during the observations.  We thank the editor and the anonymous referee for their careful review and very positive feedback. Davide Gandolfi thanks Gabriele Cologna for the interesting conversations on the properties of the planetary system. Hannu Parviainen has received support from the Rocky Planets Around Cool Stars (RoPACS) project during this research, a Marie Curie Initial Training Network funded by the European Commission's Seventh Framework Programme. He has also received funding from the V\"ais\"al\"a Foundation through the Finnish Academy of Science and Letters and from the Leverhulme Research Project grant RPG-2012-661. Financial supports from the Spanish Ministry of Economy and Competitiveness (MINECO) are acknowledged by Hans J. Deeg for the grant AYA2012-39346-C02-02, by Sergio Hoyer for the 2011 \emph{Severo Ochoa} program SEV-2011-0187, and Roi Alonso for the Ram\'on y Cajal program RYC-2010-06519. This paper includes data collected by the \kepler\ mission. Funding for the \kepler\ mission is provided by the NASA Science Mission Directorate. The \kepler\ data presented in this paper were obtained from the Mikulski Archive for Space Telescopes (MAST). STScI is operated by the Association of Universities for Research in Astronomy, Inc., under NASA contract NAS5-26555. Support for MAST for non-HST data is provided by the NASA Office of Space Science via grant NNX13AC07G and by other grants and contracts. This research has made an intensive use of the Simbad database and the VizieR catalogue access tool, CDS, Strasbourg, France. The original description of the VizieR service was published in \aaps, 143, 23.
\end{acknowledgements}


\begin{table*}
\centering
\caption{Kepler-423 system parameters.}            
\begin{tabular}{l r}
\hline
\hline
\noalign{\smallskip}
\multicolumn{2}{l}{\emph{Model parameters}} \\
\noalign{\smallskip}
\hline
\noalign{\smallskip}
Planet orbital period \pper (day)                                     &  $2.68432850\pm0.00000007$  \\
\noalign{\smallskip}
Planetary mid-transit epoch\tablefootmark{a} \ptc (BJD$_\mathrm{TDB}$-2\,450\,000 day) & $4966.35480997\pm0.00002124$\\
\noalign{\smallskip}
Planet-to-star area ratio \paa                                        &     $0.015872\pm0.000062$   \\
\noalign{\smallskip}
Planet-to-star surface brightness ratio \pfr                          & $(8.93\pm4.13) \times 10^{-4}$ \\
\noalign{\smallskip}
Impact parameter $b$                                                  &      $0.3006\pm0.0100$      \\
\noalign{\smallskip}
Bulk stellar density \srho (\gcm3)                                    &       $1.398\pm0.096$       \\
\noalign{\smallskip}
Linear limb-darkening coefficient $u_1$                               &      $0.4650\pm0.0100$      \\
\noalign{\smallskip}
Quadratic limb-darkening coefficient $u_2$                            &      $0.1518\pm0.0228$      \\
\noalign{\smallskip}
Radial velocity semi-amplitude $K$ (\kms)                             &       $0.0967\pm0.0118$     \\
\noalign{\smallskip}
Systemic radial velocity $V_\gamma$ (\kms)                            &     $-3.0410\pm0.0081$      \\
\noalign{\smallskip}
Orbit eccentricity $e$                                                &  $0.019^{+0.028}_{-0.014}$  \\
\noalign{\smallskip}
Argument of periastron $\omega$ (degree)                              & $120.26^{+77.01}_{-33.88}$  \\                             
\noalign{\smallskip}
\kepler\ LC data scatter $\sigma_\mathrm{LC}$ (ppm)                   &       $292.2\pm2.4$         \\
\noalign{\smallskip}
\kepler\ SC data scatter $\sigma_\mathrm{SC}$ (ppm)                   &      $1145.7\pm2.1$         \\
\noalign{\medskip}
\multicolumn{2}{l}{\emph{Derived parameters}} \\
\noalign{\smallskip}
\hline
\noalign{\smallskip}
Planet-to-star radius ratio $R_\mathrm{p}/R_{\star}$                   & $0.12599\pm0.00024$       \\
\noalign{\smallskip}
Planetary eclipse depth $\Delta F_{\mathrm{ec}}$ (ppm)                 & $14.2\pm6.6$              \\
\noalign{\smallskip}
Scaled semi-major axis of the planetary orbit $a_\mathrm{p}/R_{\star}$ & $8.106^{+0.117}_{-0.259}$ \\
\noalign{\smallskip}
Semi-major axis of the planetary orbit $a_\mathrm{p}$ (AU)             & $0.03585^{+0.00052}_{-0.00114}$  \\
\noalign{\smallskip}
Orbital inclination angle $i_\mathrm{p}$ (degree)                      &       $87.828\pm0.126$     \\
\noalign{\smallskip}
Planetary transit duration $T_{14}$ (hour)                             &       $2.7220\pm0.0019$     \\
\noalign{\smallskip}
Transit ingress and egress duration $T_{12}=T_{34}$ (hour)             &       $0.3330\pm0.0024$     \\
\noalign{\medskip}
\multicolumn{2}{l}{\emph{Stellar fundamental parameters}} \\
\noalign{\smallskip}
\hline
\noalign{\smallskip}
Effective temperature $T_\mathrm{eff}$ (K)                             &   $5560\pm80$        \\
\noalign{\smallskip}
Surface gravity\tablefootmark{b} log\,$g$ ~~(log$_{10}$ \cms2)        &   $4.44\pm0.10$      \\
\noalign{\smallskip}
Surface gravity\tablefootmark{c} log\,$g$ ~~(log$_{10}$ \cms2)        &   $4.41\pm0.04$      \\
\noalign{\smallskip}
Metallicity $[\mathrm{M/H}]$ (dex)                                     &   $-0.10\pm0.05$      \\
\noalign{\smallskip}
Microturbulent velocity\tablefootmark{d} $v_ {\mathrm{micro}}$ (\kms)  &    $1.0\pm0.1$       \\
\noalign{\smallskip}
Macroturbulent velocity\tablefootmark{d} $v_ {\mathrm{macro}}$ (\kms)  &    $2.8\pm0.4$       \\
\noalign{\smallskip}
Projected stellar rotational velocity \vsini\ (\kms)                   &    $2.5\pm0.5$       \\
\noalign{\smallskip}
Spectral type\tablefootmark{e}                                         &       G4\,V          \\
\noalign{\smallskip}
Star mass $M_{\star}$ (\Msun)                                          &   $0.85\pm0.04$      \\
\noalign{\smallskip}
Star radius $R_{\star}$ (\Rsun)                                        &   $0.95\pm0.04$      \\
\noalign{\smallskip}
Star age $t$ (Gyr)                                                     &     $11\pm2$         \\ 
\noalign{\smallskip}
Star rotation period\tablefootmark{f} $P_\mathrm{rot}$ (day)           &   $22.047\pm0.121$   \\
\noalign{\smallskip}
Interstellar extinction $A_\mathrm{V}$ (mag)                           &   $0.044\pm0.044$    \\
\noalign{\smallskip}
Distance of the system $d$ (pc)                                        &    $725\pm75$        \\
\noalign{\medskip}
\multicolumn{2}{l}{\emph{Planetary fundamental parameters}} \\
\noalign{\smallskip}
\hline
\noalign{\smallskip}
Planet mass\tablefootmark{g} $M_\mathrm{p}$ (\Mjup)                    &    $0.595\pm0.081$   \\
\noalign{\smallskip}
Planet radius\tablefootmark{g} $R_\mathrm{p}$ (\Rjup)                  &    $1.192\pm0.052$   \\
\noalign{\smallskip}
Planet density $\rho_\mathrm{p}$ (\gcm3)                               &    $0.459\pm0.083$   \\
\noalign{\smallskip}
Equilibrium temperature $T_\mathrm{eq}$ (K)                            &     $1605\pm120$     \\
\noalign{\smallskip}
Brightness temperature\tablefootmark{h} $T_\mathrm{br}$ (K)            &     $1950\pm250$     \\
\noalign{\smallskip}
Geometric albedo\tablefootmark{h} $A_\mathrm{g}$                       &    $0.055\pm0.028$   \\ 
\noalign{\smallskip}
Bond albedo\tablefootmark{h} $A_\mathrm{B}$                            &    $0.037\pm0.019$   \\ 
\noalign{\smallskip}
\hline       
\end{tabular}
\tablefoot{~\\
  \tablefoottext{a}{BJD$_\mathrm{TDB}$ is the barycentric Julian date in barycentric dynamical time.}~\
  \tablefoottext{b}{Obtained from the spectroscopic analysis.}~\ 
  \tablefoottext{c}{Obtained from $T_\mathrm{eff}$, $[\mathrm{M/H}]$, and $\rho_{\star}$, along with the Pisa Stellar Evolution Data Base for low-mass stars.}~\
  \tablefoottext{f}{Using the calibration equations of \citet{Bruntt2010} and \citet{Doyle2014}.}~\
  \tablefoottext{e}{With an accuracy of $\pm\,1$ sub-class.}~\
  \tablefoottext{f}{From \citet{McQuillan2013}.}~\
  \tablefoottext{g}{Radius and mass of Jupiter taken as 6.9911$\times10^9$~cm and 1.89896$\times$10$^{30}$~g, respectively.}~\
  \tablefoottext{h}{Assuming $A_\mathrm{g}=1.5 \times A_\mathrm{B}$ and heat redistribution factor between 1/4 and 2/3.}
}
\label{Parameter-Table}  
\end{table*}

~\\
\emph{\noteaddname}: \citet{Endl2014} presented an independent spectroscopic confirmation of the planetary nature of Kepler-423b. While their estimate of the planet radius (\rp~$=1.200\pm0.065$~\Rjup) agrees very well with ours, their planetary mass of \mp~$=0.72\pm0.12$~\Mjup\ is slightly ($\sim$1-$\sigma$) higher than our value. This is mainly due to their higher estimate of the stellar mass ($M_{\star}=1.07\pm0.05$~\Msun), which in turn results from an hotter stellar effective temperature (\teff~$=5790\pm116$~K) and higher iron content ([Fe/H]~$=0.26\pm0.12$~dex).

\end{document}